\documentclass[twocolumn]{aastex62}

\usepackage{gensymb}
\usepackage{amsmath}

\newcommand{\mmol}{{$M_\mathrm{H_2}$}}

\newcommand{\lco}{$L^\prime_\mathrm{CO\mbox{\scriptsize (1--0)}}$}
\newcommand{\lir}{$L_\mathrm{IR}$}

\newcommand{\aco}{$\alpha_\mathrm{CO}$}
\newcommand{\uaco}{\mbox{$M_\odot\,\mathrm{(K\,km\,s^{-1}\,pc^{2})^{-1}}$}}
\newcommand{\comol}{{CO-to-H$_2$}}
\newcommand{\umean}{{$\langle U \rangle$}}
\font\sevenrm=cmr7 scaled 1000

\newcommand{\tnm}{\tablenotemark}
\newcommand{\mcl}{\multicolumn}
\newcommand{\ph}{\phantom}
\newcommand{\phtf}{\phantom{\hspace{-5.pt}}}
\newcommand{\linmix}{\texttt{Linmix}}
\newcommand{\galfit}{\texttt{GALFIT}}
\newcommand{\CII}{[C{\sevenrm\,II}]}
\newcommand{\kms}{km\,s$^{-1}$}
 
 \font\sevenrm=cmr7 scaled 1000

\graphicspath{{./}{figures/}}

\shorttitle{CO(2--1) Survey of Quasars}
\shortauthors{Shangguan et al.}

\begin{document}

\title{An ALMA CO(2--1) Survey of Nearby Palomar-Green Quasars}

\correspondingauthor{Jinyi Shangguan}
\email{shangguan@mpe.mpg.de}

\author[0000-0002-4569-9009]{Jinyi Shangguan}
\affil{Max-Planck-Institut f\"{u}r Extraterrestrische Physik (MPE), 
Giessenbachstr., D-85748 Garching, Germany}
\affiliation{Kavli Institute for Astronomy and Astrophysics, Peking University,
Beijing 100871, China}

\author[0000-0001-6947-5846]{Luis C. Ho}
\affil{Kavli Institute for Astronomy and Astrophysics, Peking University,
Beijing 100871, China}
\affiliation{Department of Astronomy, School of Physics, Peking University,
Beijing 100871, China}

\author[0000-0002-8686-8737]{Franz E. Bauer}
\affil{Instituto de Astrof{\'{\i}}sica and Centro de Astroingenier{\'{\i}}a, Facultad 
de F{\'{i}}sica, Pontificia Universidad Cat{\'{o}}lica de Chile, Casilla 306, 
Santiago 22, Chile} 
\affiliation{Millennium Institute of Astrophysics (MAS), Nuncio Monse{\~{n}}or 
S{\'{o}}tero Sanz 100, Providencia, Santiago, Chile} 
\affiliation{Space Science Institute, 4750 Walnut Street, Suite 205, Boulder, 
Colorado 80301}

\author[0000-0003-4956-5742]{Ran Wang}
\affil{Kavli Institute for Astronomy and Astrophysics, Peking University,
Beijing 100871, China}
\affiliation{Department of Astronomy, School of Physics, Peking University,
Beijing 100871, China}

\author[0000-0001-7568-6412]{Ezequiel Treister}
\affil{Instituto de Astrof{\'{\i}}sica and Centro de Astroingenier{\'{\i}}a, Facultad 
de F{\'{i}}sica, Pontificia Universidad Cat{\'{o}}lica de Chile, Casilla 306, 
Santiago 22, Chile}



\begin{abstract}

The properties of the molecular gas can shed light on the physical conditions of 
quasar host galaxies and the effect of feedback from accreting supermassive 
black holes.  We present a new CO(2--1) survey of 23 $z < 0.1$ Palomar-Green 
quasars conducted with the Atacama Large Millimeter/submillimeter Array.  CO 
emission was successfully detected in 91\% (21/23) of the objects, from which 
we derive CO luminosities, molecular gas masses, and velocity line widths.  
Together with CO(1--0) measurements in the literature for 32 quasars 
(detection rate 53\%), there are 15 quasars with both CO(1--0) and CO(2--1) 
measurements and in total 40 sources with CO measurements.  We find that the 
line ratio $R_{21} \equiv L^\prime_{\mbox{\scriptsize CO(2--1)}}/
L^\prime_{\mbox{\scriptsize CO(1--0)}}$ is subthermal, broadly consistent 
with nearby galaxies and other quasars previously studied.  No clear 
correlation is found between $R_{21}$ and the intensity of 
the interstellar radiation field or the luminosity of the active nucleus.  As 
with the general galaxy population, quasar host galaxies exhibit a strong, 
tight, linear \lir--\lco\ relation, with a normalization consistent with that 
of starburst systems.  We investigate the molecular-to-total gas mass fraction 
with the aid of total gas masses inferred from dust masses previously derived 
from infrared observations.  Although the scatter is considerable, the current 
data do not suggest that the \comol\ conversion factor of quasar host galaxies 
significantly differs from that of normal star-forming galaxies.

\end{abstract}


\section{Introduction} 
\label{sec:intro}

Molecular gas is a fundamental ingredient of the cold interstellar medium of 
galaxies, one that directly fuels star formation \citep{Kennicutt1998ApJ,
Bigiel2008AJ} and accretion onto supermassive black holes (BHs; 
\citealt{GarciaBurillo2012JPhCS,Combes2019AA,StorchiBergmann2019NatAs}
and references therein).  It is also a direct victim 
of the putative process of energy feedback from active galactic nuclei (AGNs; 
\citealt{Fabian2012ARAA}).  The properties of the molecular gas, therefore, are 
crucial to understand the coevolution of galaxies and their central BHs 
\citep{Kormendy2013ARAA, Heckman2014ARAA}.

The cold interstellar medium of inactive, star-forming galaxies has been 
comprehensively investigated, both for the nearby \citep{Saintonge2011MNRASa,
Saintonge2017ApJS} and distant \citep{Scoville2016ApJ,Tacconi2018ApJ} Universe.
These studies have established empirical scaling relations among basic physical
quantities, including gas mass, stellar mass, and star formation rate.  Systematic 
studies of the cold gas in AGNs are still rare, particularly for objects luminous 
enough to be considered \textit{bona fide} quasars.\footnote{Following 
historical practice \citep{Schmidt1983ApJ}, we consider AGNs with 
$M_B < -23$ mag as quasars, regardless of their radio-loudness.}  Sensitivity 
limitations compelled early investigations to focus mainly on nearby AGNs or 
mostly quasars with strong far-infrared (IR) emission.  While the molecular gas
mass of nearby Seyfert galaxies is similar to that of star-forming galaxies 
(matched in Hubble type and $B$-band luminosity), the star formation efficiency, 
as inferred from their extended far-IR emission, is higher among the Seyferts 
\citep{Maiolino1997ApJ}.  Similarly, the \lir--\lco\ relation of nearby quasars lies 
well above that of star-forming galaxies, which may indicate that the dust is heated 
by the quasar in addition to young stars \citep{Evans2001AJ,Evans2006AJ}.  
Irrespective of the detailed properties of the molecular medium, the existing 
data, scant though they may be, suggest that low-redshift, optically selected 
quasars reside in gas-rich host galaxies  \citep{Scoville2003ApJ}.  

Gas outflows, in molecular and other forms, have been observed in 
nearby AGNs (e.g., \citealt{Cicone2014AA,Feruglio2015AA,Harrison2018NatAs,
HerreraCamus2019ApJ}) and higher redshift ($z \gtrsim 1$) quasars (e.g., 
\citealt{Maiolino2012MNRAS,Bischetti2017AA,Brusa2018AA,Bischetti2019arXiv}),
plausibly interpreted as evidence of energy injection by so-called quasar-mode 
AGN feedback \citep{DiMatteo2005Natur,Hopkins2008ApJS,Fabian2012ARAA}.  There 
is no consensus, however, as to whether AGN-driven outflows truly influence the 
cold gas content of AGN host galaxies (\citealt{Ho2008ApJ, CanoDiaz2012AA,
Maiolino2012MNRAS,Cresci2015ApJ,Carniani2016AA,Bischetti2017AA,Vayner2017ApJ,
Baron2018MNRAS,Brusa2018AA,Ellison2018MNRAS,Perna2018AA,Shangguan2018ApJ,
Shangguan2019ApJ,Russell2019arXiv}).  High-redshift quasars are routinely 
detected with submillimeter tracers such as CO and \CII\ 158 $\mu$m, furnishing 
fundamental properties of their host galaxies (e.g., gas masses and dynamical 
masses) that would otherwise be inaccessible (\citealt{Walter2004ApJ,
Wang2013ApJ,Wang2016ApJ,Shao2017ApJ}).  

In this context, a comprehensive study of the local counterparts of 
high-redshift systems provides valuable insights into the coevolution of BHs 
and galaxies over cosmic time.  Key questions still linger as to whether and 
how quasars affect the cold interstellar medium of their host galaxies.  Are 
the basic properties of cold gas in quasar host galaxies different from those 
of inactive galaxies?  Are there physical links between the properties of the active 
nuclei and the cold gas on large scales?  Does star formation operate in the same 
manner as ordinary star-forming galaxies?   Using CO(1--0) and CO(2--1) observations 
of a sample of 14 nearby quasars, \cite{Husemann2017MNRAS} concluded that gas 
fraction and star formation efficiency depend on the host galaxy morphology.  
Gas fractions and gas depletion time scales in disk-dominated hosts resemble 
those of star-forming galaxies; bulge-dominated hosts, while generally more
gas-poor, appear to exhibit higher star formation efficiencies.  AGN power 
correlates strongly with molecular gas mass, pointing to a plausible causal 
link between the two, but the overall gas content of the host galaxies does 
not appear to be depleted by quasar-mode feedback.

To enlarge the sample of nearby quasars with molecular gas measurements, we 
used the Atacama Large Millimeter/submillimeter Array (ALMA) to conduct a CO 
survey of a well-defined sample of 23 low-redshift ($z < 0.1$ and 
$\mathrm{declination}<30\degree$) quasars selected from the Palomar-Green (PG; 
\citealt{Schmidt1983ApJ}) survey.  It is important to recognize that the PG sample 
was originally ultraviolet-selected, and hence was not selected based on 
the dust or gas properties of the quasars.  The high sensitivity of ALMA enabled 
us to detect CO(2--1) emission in 21 out of the 23 quasars, nearly 
doubling the number of CO detections of PG quasars known to date.  Combined 
with previous results from the literature, we now have measurements of either 
CO(1--0) or CO(2--1) for a representative subset of 40 out of the parent sample 
of 70 PG quasars at $z<0.3$, for which we provide self-consistent measurements 
of CO luminosity, molecular gas mass, and velocity line width.  The focus of 
this paper is to describe our sample and present basic physical quantities for it.  
A companion paper \citep{Shangguan2019bApJ} investigates the possible 
connections between the properties of the AGN and the molecular gas.

We introduce the sample and observations in Section \ref{sec:obs}.  The methods 
to derive the physical quantities are described in Section \ref{sec:mea}.  Section 
\ref{sec:dis} discusses the CO line ratio, the \lir--\lco\ relation of quasar host galaxies, 
and the \comol\ conversion factor.  We adopt the following cosmological parameters: 
$\Omega_m = 0.308$, $\Omega_\Lambda = 0.692$, and $H_{0}=67.8$ \kms\ Mpc$^{-1}$ 
\citep{Planck2016AA}.

\section{Sample and Observations} 
\label{sec:obs}

\begin{figure*}
\begin{center}
\includegraphics[width=0.95\textwidth]{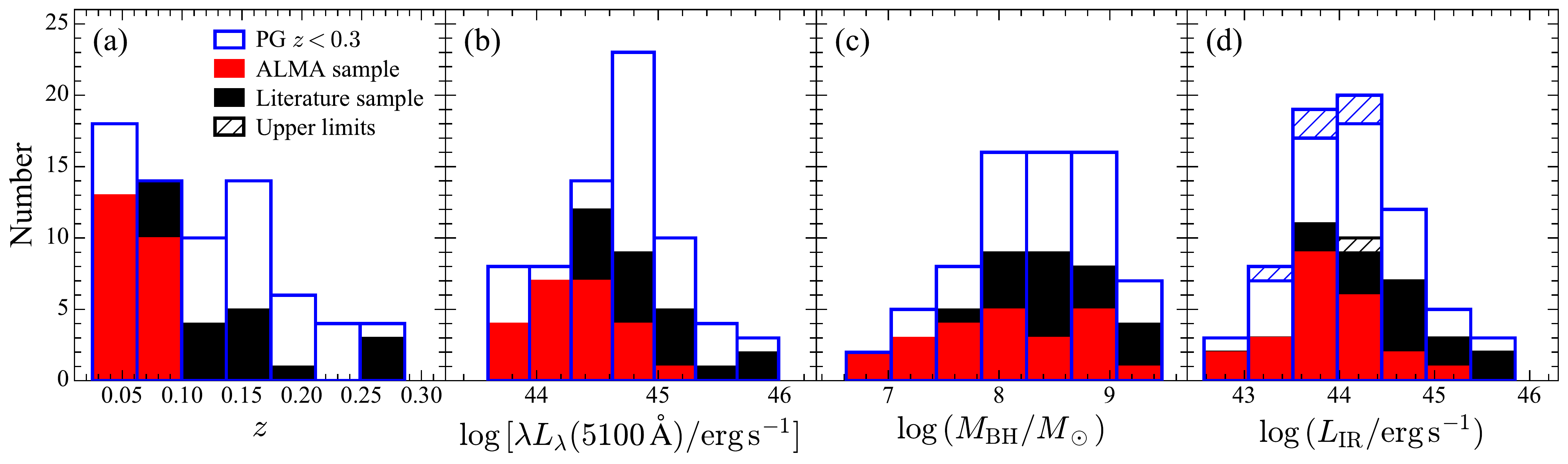}
\caption{Comparison of the parent sample of 70 $z<0.3$ PG quasars (blue) 
with the subsample of 40 PG quasars with CO measurements from our ALMA 
observations (red; 23 sources) and from the literature (black; 13 sources), 
in terms of (a) redshift, (b) 5100 \AA\ AGN luminosity, (c) BH mass, and (d) 
total IR (8--1000 \micron) luminosity.  The hatched areas indicate the objects
with \lir\ upper limits.}
\label{fig:smp} 
\end{center}
\end{figure*}

Our sample derives from the lower redshift subset of the 
ultraviolet/optically selected quasars from the PG survey.  PG quasars have 
been extensively studied for decades, allowing us to take advantage of a wealth 
of available multiwavelength data.  \cite{Shangguan2018ApJ} performed a 
comprehensive analysis of the IR spectral energy distributions of the 87 PG 
quasars with $z < 0.5$ \citep{Boroson1992ApJS} to derive robust dust masses 
and total IR (8--1000 \micron) luminosities of the host galaxies.  They used 
the dust masses to infer global total (atomic plus molecular) gas masses.  We 
directly use the 5100 \AA\ AGN continuum luminosities, BH masses, and host 
galaxy stellar masses compiled by them.  The stellar masses are derived from 
high-resolution optical/near-IR images with the nuclear emission decomposed 
\citep{Zhang2016ApJ}.\footnote{The stellar mass is not available from the 
decomposition of integrated spectral energy distribution, mainly due to the 
contamination of the overwhelming nuclear emission.}  For objects without 
stellar mass measurements, \cite{Shangguan2018ApJ} used bulge masses estimated 
from the $M_\mathrm{BH}$--$M_\mathrm{bulge}$ relation \citep{Kormendy2013ARAA}.  
The axis ratio ($q$) of the host galaxy comes from two-dimensional \galfit\ 
\citep{Peng2002AJ,Peng2010AJ} decomposition of high-resolution optical and 
near-IR images acquired with the \textit{Hubble Space Telescope} 
(\citealt{Kim2017ApJS}; Y. Zhao et al., in preparation).

We observed the $^{12}$CO(2--1) 230.538 GHz line for all 23 PG quasars with 
$z<0.1$ using the Band-6 receiver of the ALMA Compact Array (ACA)  during 
Cycle 5 (PI: F. Bauer, 103.1 hours in total).  The brighness of these nearby 
quasars allows to obtain significant detections using ACA with moderate resolving 
power (FWHM $\approx 6\arcsec$) in relatively short exposure times. 
Table~\ref{tab:obs} gives a summary of the 
observations.  The flux and bandpass calibrators were observed in the beginning 
of each observation, and the phase calibrator was observed every $\sim 5-10$ 
minutes.  The on-source integration times lasted between 120 and 280 min, 
typically 150 min.  Integration times were estimated based on the CO(2--1) 
brightness expected from the AGN-decomposed IR luminosity of 
\cite{Shangguan2018ApJ}, assuming that the \lir--\lco\ relation is given by 
Equation (1) of \cite{Sargent2014ApJ} for starburst galaxies, adopting a CO 
line luminosity ratio 
$R_{21}\equiv L^\prime_\mathrm{CO(2-1)}/L^\prime_\mathrm{CO(1-0)} = 0.5$ 
\citep{Xia2012ApJ}.  The data cube covers $\gtrsim 4000$ \kms, spanning 
the full $\sim 3.6$ GHz spectral window of one sideband.  The Hanning-smoothed 
spectral resolution is $\sim 5$ \kms.  

We reduced the data with the Common Astronomy Software 
Application\footnote{\url{https://casa.nrao.edu}} (CASA; \citealt{McMullin2007ASP}).  
The data were calibrated using the standard pipeline after minor flaggings of 
some problematic antennae and channels with sky absorption lines; this process
did not affect the final results significantly.  The continuum is 
subtracted with \texttt{uvcontsub}, fitting channels away from the line 
emission.  Line images were constructed using the task \texttt{CLEAN} with 
robust weighting ({\tt robust = 0.5}) and a stop threshold 2.5 times the root 
mean square (rms) of the off-source channels.  The measured 1 $\sigma$ 
noise level per beam per channel (typically 1.5--5 mJy) is consistent 
within $\sim 30 \%$ of the theoretical noise limit.  To confirm that 
the ACA robust beam recovers all of the flux, we extracted fluxes using a
15\arcsec\ tapered beam, finding a 1:1 ratio within 3 $\sigma$ for all objects 
(Table \ref{tab:alma}). 

Together with published $^{12}$CO(1--0) data for 17 additional objects, there are 
CO measurements for 40 $z<0.3$  PG quasars.  Figure \ref{fig:smp} compares the 
CO-measured subsample with the parent sample of 70 PG quasars with $z<0.3$.
Although the redshift distribution of the CO-measured objects is dominated by 
objects at $z \lesssim 0.1$, a two-sample Peto-Prentice test\footnote{The 
Peto-Prentice test is adopted to work with samples including censored data (in 
our case, IR luminosity).  It is equivalent to the Gehan test when there are 
no censored data.} finds that the two redshift distributions are not 
statistically different; the probability of the null hypothesis that 
the distributions are drawn from the same parent sample is $P_{\rm null} = 
10.4\%$.  The same holds for the distributions of 5100 \AA\ AGN luminosity, BH 
mass, and IR luminosity, for which $P_{\rm null} = 37.6\%$, 58.7\%, and 71.3\%, 
respectively.  We conclude that the CO-measured sample is representative of the 
parent sample of $z<0.3$ PG quasars.  

\section{Measurements}
\label{sec:mea}

\subsection{CO Luminosity}
\label{ssec:lco}

\begin{figure*}
\begin{center}
\includegraphics[height=0.3\textheight]{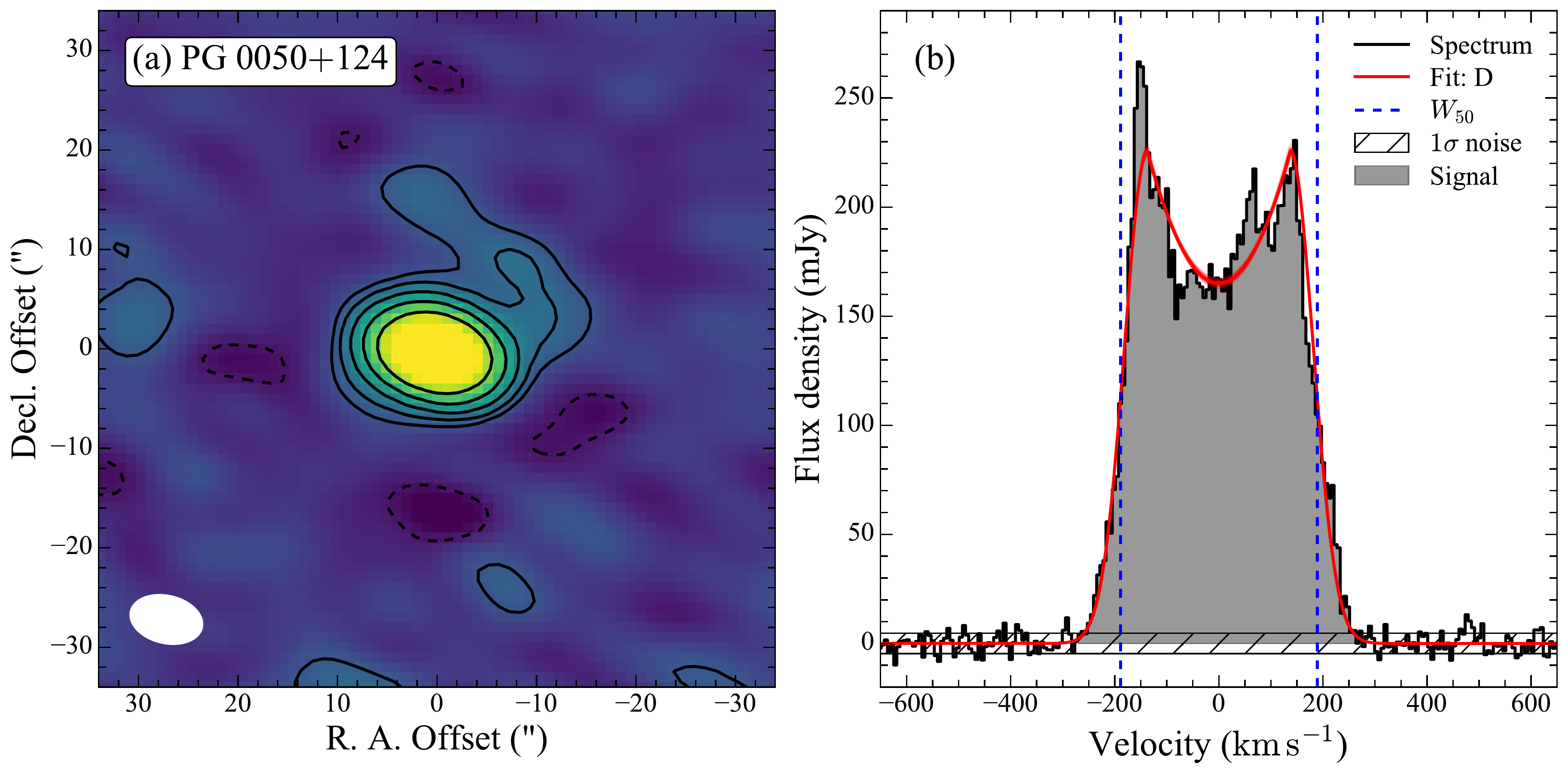}
\caption{(a) CO(2--1) intensity (moment~0) map of PG 0050+124.  The contours 
are $-$2 (dashed), 2, 4, 8, 16, and 32 $\sigma$ levels, with $\sigma$ being the 
rms of the source-free pixels in the map.  The synthesis 
beam is indicated on the lower-left corner of the map.  The beam is $7\farcs4 
\times 4\farcs8$ with a position angle of 77\degree.  (b) The one-dimensional 
spectrum extracted from the 2 $\sigma$ contour of the source emission.  
Channels shaded in grey are considered to be signal from the emission line.  
The hatched horizontal band indicates the noise level of the line-free channels.
The emission line is fit with a double-peak Gaussian profile (red curve).  The 
full width of the 50 percentile of the best-fit profile, $W_{50}$, is indicated by 
the blue dashed lines.}
\label{fig:drex} 
\end{center}
\end{figure*}

We use the channels above the 1 $\sigma$ level of the spectrum to generate 
the CO intensity (moment~0) map (Figure \ref{fig:drex}a). The integrated CO 
flux is measured from the intensity map by summing up the pixels within the 
2 $\sigma$ contour of the source emission.  We estimate the uncertainty from 
the standard deviation of 20 repeated off-source measurements using 
a circular aperture containing the same number of pixels as those within the 
2 $\sigma$ contour of the source.  The uncertainty of the absolute flux scale, 
$\sim 5\%-10\%$ \citep{Fomalont2014Msngr, Bonato2018MNRAS}, is not 
included in our final flux uncertainty.  The integrated CO spectrum, used to 
measure the line width (Section \ref{ssec:lw}), is extracted from the line-emitting 
region above 2 $\sigma$ of the intensity map.  CO(2--1) was previously 
detected in PG~0050+124 using the JCMT (114$\pm$23 Jy km s$^{-1}$; 
\citealt{Papadopoulos2008AA}) and in PG~1126$-$041 using IRAM 30~m 
(24.7$\pm$1.6 Jy km s$^{-1}$; \citealt{Bertram2007AA}).\footnote{S/T=8.19 
Jy/K is assumed for the IRAM 30~m measurement.}  Our line fluxes are 
reasonably consistent, with deviations $\lesssim 50\%$.

Following \cite{Solomon2005ARAA}, the CO line luminosity is
\begin{equation}
L^\prime_\mathrm{CO} = 3.25 \times 10^7\, S_\mathrm{CO}\Delta\nu\,
\nu_\mathrm{obs}^{-2}\, D_L^2\, (1+z)^{-3},
\end{equation}

\noindent 
where $L^\prime_\mathrm{CO}$ is the CO line luminosity in units of 
$\mathrm{K\,km\,s^{-1}\,pc^2}$, $S_\mathrm{CO}\Delta\nu$ is the integrated 
line flux in units of $\mathrm{Jy\,km\,s^{-1}}$, $\nu_\mathrm{obs}$ is the 
observed frequency of the CO(2--1) line in GHz, and $D_L$ is the luminosity 
distance in Mpc.  The factor \aco\ is needed to derive molecular gas masses 
from the CO luminosity (\citealt{Bolatto2013ARAA}, and references therein).  
Since \aco\ is usually quoted for the CO(1--0) line, we need to convert the 
line luminosity from $L_{\mbox{\scriptsize CO(2--1)}}$ to 
$L_{\mbox{\scriptsize CO(1--0)}}$ in order to derive the molecular gas 
mass.  Fortunately, literature measurements of CO(1--0) are available for 15 
of the objects in our ALMA sample, among them eight detections (Section 
\ref{ssec:lida}).  We find a median value of $R_{21} = 0.62^{+0.15}_{-0.07}$ 
(Section \ref{ssec:r21}).  We adopt this median value of $R_{21}$ to convert 
all the new ALMA CO luminosities from $L_{\mbox{\scriptsize CO(2--1)}}$ 
to $L_{\mbox{\scriptsize CO(1--0)}}$ (Table \ref{tab:alma}).  We do not 
consider the uncertainty on $R_{21}$, as it is hardly well-constrained by our data.  
However, if $R_{21}$ varies from 0.5 to 1.0, it could contribute to the final 
uncertainty of \mmol\ as significantly as \aco\ ($\sim 0.3$ dex).  It is reassuring 
that our estimated value of $R_{21}$ agrees well with values found in nearby 
galaxies and AGNs (\citealt{OcanaFlaquer2010AA,Sandstrom2013ApJ, 
Rosolowsky2015AAS,Husemann2017MNRAS,Saintonge2017ApJS}; see 
Section \ref{ssec:r21}).

\subsection{CO Line Width}
\label{ssec:lw}

The velocity width of the integrated emission-line profiles of galaxies is 
commonly specified as the line width at 20 percent ($W_{20}$; e.g., 
\citealt{Tully1977AA}) or 50 percent ($W_{50}$; e.g., \citealt{Tiley2016MNRAS})
of the peak intensity.  For spectra with relatively low signal-to-noise ratio, 
we can obtain more accurate line widths by fitting a model line profile to the
data instead of measuring them directly from the observed spectrum.  With the 
aid of a suite of integrated spectra of simulated galaxies, 
\cite{Tiley2016MNRAS} evaluated the effectiveness of various methods for fitting
line profiles and concluded that the ``double-peak'' Gaussian function---a 
parabolic function bordered by a half-Gaussian symmetrically on either 
side---provides the most robust measure of $W_{50}$.  The double-peak Gaussian 
function is defined as \citep{Tiley2016MNRAS}
\begin{equation}
  f(v) =  \left \{
  \begin{array}{ll}
  A_\mathrm{G} \times \exp{\frac{-[v - (v_0 - w)]^2]}{2\sigma^2}} & v < v_0 - w \\
  A_\mathrm{C} + a (v - v_0)^2 & v_0 - w \le v \le v_0 + w \\
  A_\mathrm{G} \times \exp{\frac{-[v - (v_0 + w)]^2]}{2\sigma^2}} & v > v_0 - w
  \end{array}
  \right.,
\end{equation}
\noindent
where $-500\,{\rm km\,s^{-1}} < v_0 < 500\,{\rm km\,s^{-1}}$ is the central 
velocity, $w$ ($>0\,{\rm km\,s^{-1}}$) is the half width of the central 
parabola, $\sigma$ ($>0\,{\rm km\,s^{-1}}$) is the width of the edge 
half-Gaussian profile, $A_\mathrm{G}>0$ is the peak flux of the half-Gaussian 
edges at $v_0 \pm w$, $A_\mathrm{C}$ is the flux at the profile center, and 
$a=(A_\mathrm{G}-A_\mathrm{C})/w^2$.  Then, the two conventionally used line 
widths are given by
\begin{equation}\label{eq:lw}
\begin{aligned}
W_{50}=2(w+\sqrt{2\ln 2} \sigma), \\
W_{20}=2(w+\sqrt{2\ln 5} \sigma).
\end{aligned}
\end{equation}
\noindent
Equation (\ref{eq:lw}) cannot accurately specify the width of strongly convex, 
single-peaked profiles.  Under these circumstances, the data should be fit with 
a standard Gaussian function,\footnote{The Gaussian function is simply $f(v)=
A\exp{\frac{-(v-v_0)^2}{2\sigma^2}}$.} for which $W_{50}=2\sqrt{2\ln2}\sigma$ and
$W_{20}=2\sqrt{2\ln5}\sigma$.  Following \cite{Tiley2016MNRAS}, we adopt the 
standard Gaussian function when either of these two criteria holds: (1) the 
reduced chi-square of the standard Gaussian fit is closer to unity than that 
of the double-peak Gaussian function;\footnote{The reduced $\chi^2$ is defined 
as $\left(\sum_i \frac{[F(v_i)-f(v_i)]^2}{\sigma_\mathrm{rms}^2}\right)/N$, 
where $F(v_i)$ is the observed CO(2--1) spectral flux density in velocity bin 
$v_i$, $f(v_i)$ is the model flux density, $\sigma_\mathrm{rms}$ is the rms 
noise from the line-free channels, $N$ is the number of degrees of freedom, 
and the sum is taken over all of the channels.} (2) $A_\mathrm{G}/A_\mathrm{C} 
< 2/3$.  We use a Markov chain Monte Carlo method in the \texttt{emcee} package 
\citep{ForemanMackey2013PASP} to perform the fit.

An example of the profile-fitting method is shown in Figure \ref{fig:drex}b.  
PG~0050+124 is one of the brightest objects in our sample.
Appendix \ref{apd:plt} gives the data for the remaining 20 detected objects, 
all of which were successfully fit, apart from the tentative detection of 
PG~1341+258, which suffers from exceptionally low signal-to-noise ratio.  
Table~\ref{tab:alma} lists measurements of both $W_{50}$ and $W_{20}$, 
the latter because sometimes only this quantity is reported in the 
literature; we need to use our measured $W_{20}/W_{50}$ ratios to incorporate 
the published line widths into our analysis (Section \ref{ssec:lida}).  The 
systemic velocities of the CO line agree closely ($<5\%$ difference) with the 
optical redshifts, and for our final analysis we simply adopt the latter.

\subsection{Measurements from the Literature}
\label{ssec:lida}

To date, CO(1--0) measurements have been published for 32 PG quasars, as 
summarized in \cite{Shangguan2018ApJ}.  Among them, 15 objects were included in
our ALMA program and hence now have both CO transitions observed (Table 
\ref{tab:r21}), leaving the remaining 17 that only have CO(1--0) data (Table 
\ref{tab:lite}).  The published CO fluxes were converted to luminosities according 
to our adopted cosmological parameters.  Line widths were reported as either 
$W_{20}$ or $W_{50}$, usually with no uncertainties specified.  We homogenize 
the line widths adopting $W_{20}/W_{50} = 1.17 \pm 0.19$, the median 
ratio measured in our ALMA sample.  Of the 15 quasars with both CO(1--0) and 
CO(2--1) observations, eight are detected in both lines.\footnote{According to
\cite{Shangguan2018ApJ}, the 3 $\sigma$ CO(1--0) detections of PG~0003+199 
\citep{Maiolino1997ApJ} and PG~2214+139 \citep{Scoville2003ApJ} were likely 
overestimated.  We regard them as upper limits.}  These objects 
provide valuable insight on $R_{21}$ (Section \ref{ssec:r21}), which is 
needed to convert the line luminosity from CO(2--1) to CO(1--0), as discussed 
in Section \ref{ssec:lco}.

\section{Discussion} 
\label{sec:dis}

\subsection{The CO(2--1)/CO(1--0) Ratio}
\label{ssec:r21}

As listed in Table \ref{tab:r21}, the line ratios of the eight quasars in our 
study with both lines detected span $R_{21} = 0.49-0.90$. The $50^{+25}_{-25}$th 
percentile value, calculated with the Kaplan-Meier product-limit estimator 
\texttt{kmestimate} in \texttt{IRAF.ASURV} \citep{Feigelson1985ApJ,Lavalley1992ASPC}, 
is $R_{21} = 0.62^{+0.15}_{-0.07}$.  If the CO emission is thermalized and optically 
thick, the intrinsic brightness temperature and luminosity of the line are independent 
of $J$ and rest frequency, and $R_{21}=1$.  Indeed, a value of $R_{21} \approx 1$ 
is observed in the inner parts of spiral galaxies (\citealt{Braine1992AA}), 
local luminous IR galaxies \citep{Papadopoulos2012MNRAS}, and high-redshift 
galaxies (\citealt{Carilli2013ARAA,Daddi2015AA}).  However, recent studies find 
lower values of $R_{21}$ on the global scales of nearby galactic disks 
($R_{21} \lesssim 0.8$; \citealt{Leroy2013AJ,Rosolowsky2015AAS,Saintonge2017ApJS}).
\cite{OcanaFlaquer2010AA} report $R_{21} \approx 0.6$ for nearby radio galaxies,
and some low-redshift quasars can reach $R_{21} \approx 0.5$ 
\citep{Husemann2017MNRAS}, while in IR-luminous quasars $R_{21}\approx 0.4-1.2$,
with a mean value of $\sim 0.8$ \citep{Xia2012ApJ}.  Therefore, our low-redshift 
quasars exhibit $R_{21}$ values fully consistent with those derived 
from global measurements of nearby inactive and active galaxies.  In contrast, 
high-redshift quasars show low-$J$ CO line ratios suggestive of optically thick, 
thermally excited emission, indicating that the molecular gas emission comes 
from a compact region in the centers of the host galaxies 
\citep{Carilli2013ARAA}.

The relative spatial coverage of CO(1--0) and CO(2--1) introduces additional 
uncertainties into the interpretation of $R_{21}$, especially for single-beam 
observations of nearby galaxies when both lines are observed with the same 
telescope\footnote{The IRAM 30~m beam size is 22\arcsec\ for CO(1--0) and 
11\arcsec\ for CO(2--1).} or interferometer configuration.  To 
alleviate such complications, it is customary to scale down the flux of 
CO(1--0) to match that of CO(2--1), often limiting the measurement of the line 
ratio to the central part of the galaxy.  For example, \cite{Husemann2017MNRAS}
use H$\alpha$ emission to estimate the spatial distribution of CO(1--0) and 
scale down the flux of CO(1--0) to match that of CO(2--1).  
\cite{Saintonge2017ApJS}, in contrast, observe CO(2--1) using the APEX 12~m 
telescope, whose 230 GHz beam of 27\arcsec\ better matches the 22\arcsec\ 
beam of the IRAM 30~m telescope for CO(1--0).  Fortunately, the relatively 
large distances 
of our sources obviate these complications.  At $z\gtrsim0.05$, the 
CO(1--0) emission of our objects should be mostly captured by the beam 
of the IRAM 30~m telescope \citep{Evans2006AJ,Bertram2007AA}, while 
all of the CO(2--1) emission should be contained within the maximum 
recoverable scale ($\sim 29\arcsec$ at 230 GHz) of ACA, which is confirmed 
by our 15\arcsec\ tapered measurements (Section \ref{sec:obs}).   Meanwhile, 
CO(1--0) emission may still be underestimated to some extent, when the 
emission size is comparable to the beam size but its spatial distribution is 
unknown from the single-dish observation.  This may also contribute to the 
uncertainty of $R_{21}$.

\begin{figure*}
\begin{center}
\includegraphics[height=0.3\textheight]{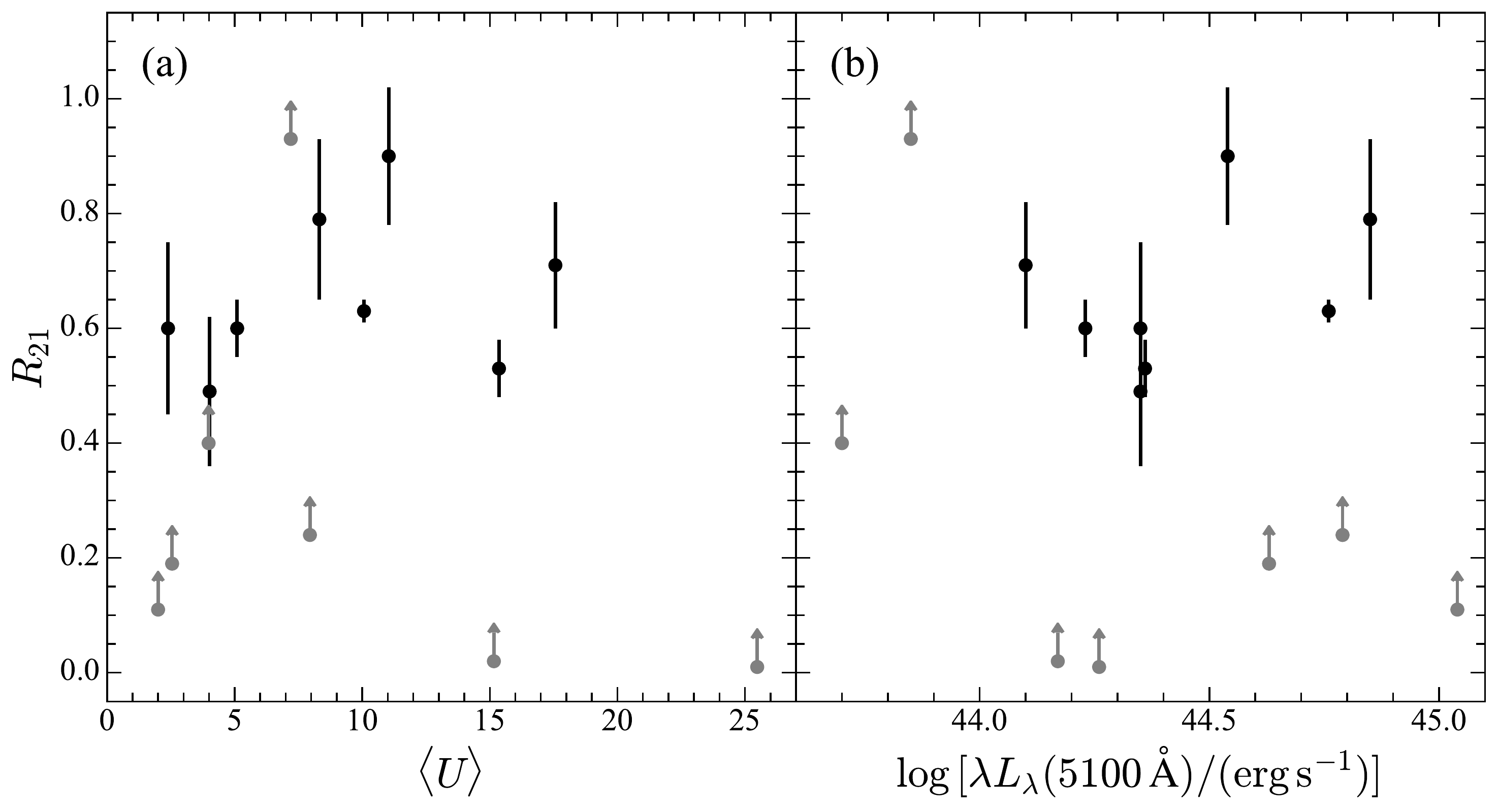}
\caption{The relation between CO line intensity ratio $R_{21}$ and (a) the 
mean intensity of the interstellar radiation field \umean\ and (b) quasar 
5100 \AA\ continuum luminosity.  Objects with CO(1--0) upper limits are 
denoted as grey symbols. A possible increasing trend is found in (a), but none 
is obvious in (b).
}
\label{fig:r21} 
\end{center}
\end{figure*}

The low values of $R_{21}$ for our quasars indicate that the molecular gas is
optically thick but either is sub-thermally excited \citep{OcanaFlaquer2010AA,
Husemann2017MNRAS} or has low temperature ($\lesssim 10$ K; 
\citealt{Braine1992AA}).  Motivated by \cite{Daddi2015AA}, who found a
significant sublinear correlation between the mean intensity of the interstellar 
radiation field of the galaxy (\umean; \citealt{Draine2007ApJ}) and the CO(5--4)/CO(2--1) 
ratio, we checked but failed to find a clear correlation between $R_{21}$ and 
\umean\ for the quasar host galaxies (Figure \ref{fig:r21}a).  Unfortunately, the number of 
objects with statistically meaningful measurements is too small to 
perform a formal statistical test.  \umean\ comes from the study of IR spectral 
energy distributions of PG quasars \citep{Shangguan2018ApJ}.  The dust 
temperatures of the quasar host galaxies, however, are $\gtrsim 20$ K 
\citep{Shangguan2018ApJ}, which are not entirely consistent with the molecular 
gas temperature of $\lesssim 10$ K expected from $R_{21} \approx 0.6$, if the 
gas is thermally excited (see Figure 1 of \citealt{Braine1992AA}).  Although a 
detailed discussion is beyond the scope of this paper, we note that the continuum 
emission of almost all of the quasars are unresolved with our ACA measurements, 
while CO(2--1) of nearly half of the quasars is resolved.\footnote{The size 
measurements are based on CASA 2D fit.} This suggests that the dust emission 
from far-IR to submillimeter is predominantly powered by an AGN or nuclear starburst.  
We do not know whether the quasar affects the excitation of the low-$J$ CO lines, 
as $R_{21}$ seems unrelated to the AGN luminosity (Figure \ref{fig:r21}b).  We 
conclude that, similar as low-$z$ galaxies and AGNs, the low-$J$ CO emission of 
our quasars is subthermally excited.

\subsection{IR versus CO Relation}
\label{ssec:lcoir}

\begin{figure}
\begin{center}
\includegraphics[height=0.3\textheight]{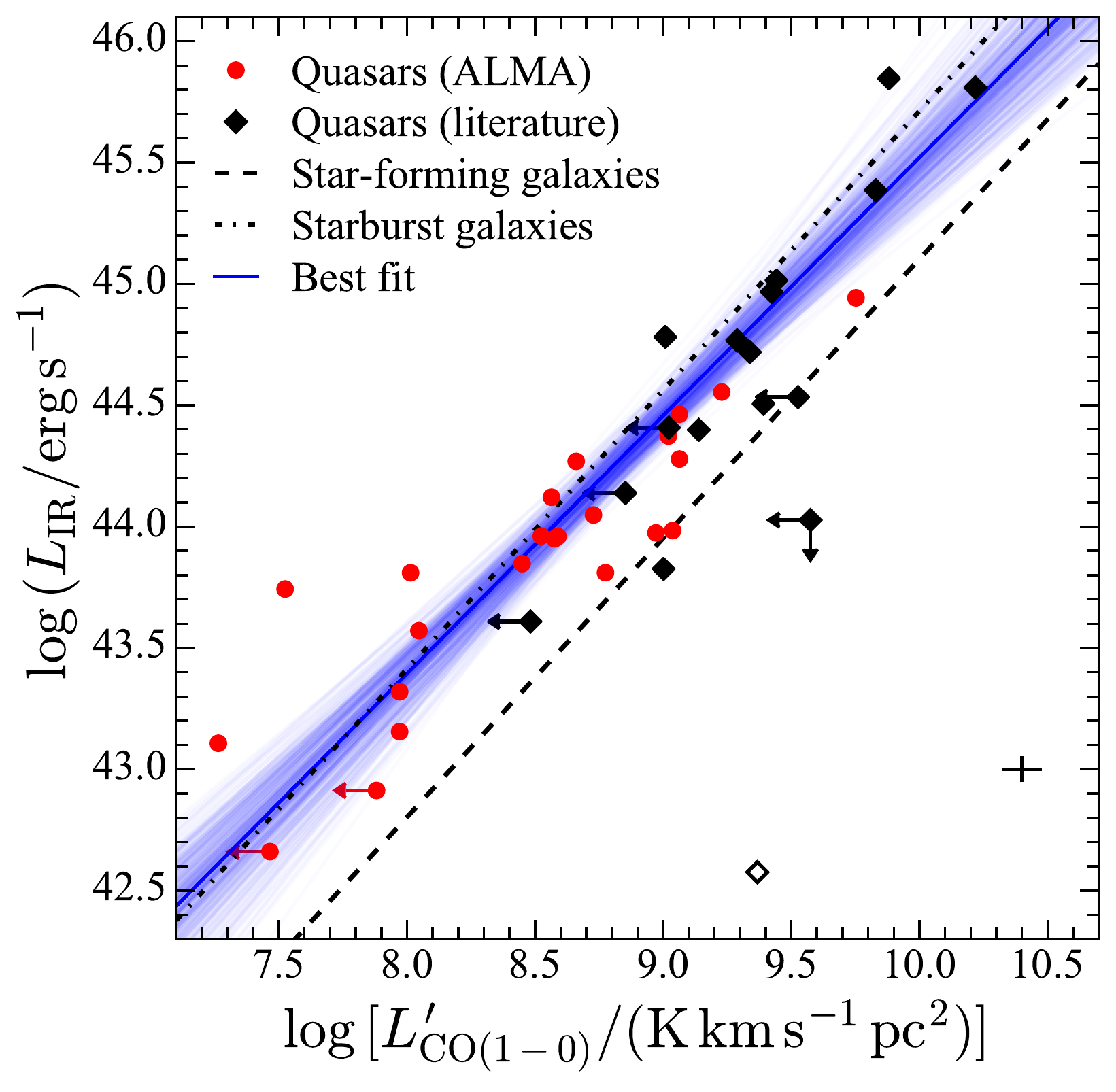}
\caption{The relation between IR and CO luminosity for quasar host galaxies,
compared with those of normal star-forming galaxies (dashed line) and nearby 
and high-redshift starburst galaxies (dash-dotted line) from \cite{Genzel2010MNRAS}.  
The blue solid line is the best-fit relation for quasars, including the upper limits 
of \lco.  The faint blue lines indicate the uncertainty of the fit.  The 90th percentiles 
of the measured uncertainties are indicated in the lower-right corner.  We 
exclude PG~1226+023 (open diamond) from the fit, as the IR luminosity of
its host galaxy is very uncertain.}
\label{fig:lco} 
\end{center}
\end{figure}

The CO line luminosity correlates strongly with the IR luminosity of galaxies, 
both active and inactive, at low and high redshifts (e.g., 
\citealt{Sanders1985ApJ,Solomon2005ARAA,Genzel2010MNRAS,Saintonge2011MNRASb,
Xia2012ApJ,Carilli2013ARAA}).  Sensitive to the Lyman continuum emission 
absorbed and reprocessed by dust \citep{Kennicutt1998ARAA}, the IR luminosity 
provides an excellent tracer of the star formation rate in star-forming 
galaxies.  Therefore, the ratio of \lir\ to \lco\ reflects the global star 
formation efficiency of the molecular gas.  In quasar host galaxies, 
however, emission from hot dust, heated by BH accretion, may dominate the IR 
luminosity and contribute a significant fraction of the emission, even up to 
$\sim 100$ \micron\ \citep{Lani2017MNRAS, Lyu2017ApJ,Zhuang2018ApJ}.  We 
calculate the $8-1000$ \micron\ IR luminosity from the cold dust emission 
decomposed from the integrated spectral energy distributions of 
\cite{Shangguan2018ApJ}.\footnote{We adopt the quantity $L_\mathrm{IR, host}$ 
from \cite{Shangguan2018ApJ}, but denote it here as \lir\ for short.}  

Figure \ref{fig:lco} compares the \lir--\lco\ relation of PG quasars with those 
of star-forming galaxies and starburst systems triggered by galaxy mergers 
\citep{Genzel2010MNRAS}.  Starburst galaxies are typically $\gtrsim 0.4$ 
dex above the so-called main sequence of the star-forming galaxies (e.g., 
\citealt{Elbaz2018AA,Shangguan2019aApJ}).  As with other types of 
galaxies, the host galaxies of quasars clearly also exhibit a strong 
correlation. We fit the relation of the PG~quasars with \linmix\ 
\citep{Kelly2007ApJ},\footnote{Since \linmix\ only allows upper limits on 
the dependent variable, we treat \lir\ as the independent variable and 
\lco\ as the dependent variable.  We do not include PG~1545+210, which 
contains upper limits in both \lir\ and \lco.  We assign an 
uncertainty of 0.1 dex to the literature measurements for which error 
estimates are unavailable, but the exact value is not critical to the fit.} 
accounting for the upper limits in \lco.  The best fit, 
\begin{equation} \label{eq:coir}
\log\,L^\prime_\mathrm{CO(1\mbox{--}0)} = 0.94\left(^{+0.08}_{-0.08}\right)\,\log\,
L_\mathrm{IR} - 32.90\left(^{+3.35}_{-3.40}\right),
\end{equation}

\noindent 
is consistent with a linear relation between \lir\ and \lco.  Both the slope 
and the zero point are consistent with the relation for starburst galaxies.  
The total scatter of the relation ($\sim 0.3$ dex) is dominated by an 
intrinsic scatter of $0.29^{+0.05}_{-0.04}$ dex.  PG~1226+023 is 
excluded from the fit because the IR luminosity of its host galaxy is very 
uncertain (see Table \ref{tab:lite} and \citealt{Shangguan2018ApJ}), but the 
fit results do not depend on this choice.  We searched for, but failed to find, a statistically 
significant partial correlation of the \lir--\lco\ relation with any plausible 
third variable (e.g., AGN luminosity).

\subsection{CO-to-H$_2$ Conversion Factor}
\label{ssec:mh2}

To the best of our knowledge, there has never been a formal study of the \comol\
conversion factor (\aco) of galaxies hosting AGNs powerful enough to qualify as 
quasars.  Here we use our new CO measurements, in combination with previous 
total gas measurements estimated from dust content \citep{Shangguan2018ApJ}, 
to put a rough constraint on \aco\ in quasar host galaxies.  As a starting point, we 
adopt \aco\ = 3.1 \uaco\ with 0.3 dex uncertainty, as recommended by 
\cite{Sandstrom2013ApJ}, who, as in \cite{Leroy2011ApJ}, simultaneously solved 
for \aco\ and the gas-to-dust ratio for 26 nearby star-forming galaxies, for the first 
time beyond the Local Group.  This value of \aco\ is slightly lower than, but consistent 
with, the canonical Milky Way value of 4.3 \uaco\ \citep{Bolatto2013ARAA}, and it 
does not appear to depend strongly on metallicity for galaxies with metallicities 
similar to and above that of the Milky Way.  While nuclear activity potentially can 
affect the molecular gas of the host (e.g., \citealt{Krips2008ApJ}), there is no clear
evidence that the presence of an AGN influences \aco\ \citep{Sandstrom2013ApJ}, 
even when AGN feedback is in principle powerful enough to be effective 
\citep{Rosario2018MNRAS}.

\begin{figure}
\begin{center}
\includegraphics[height=0.3\textheight]{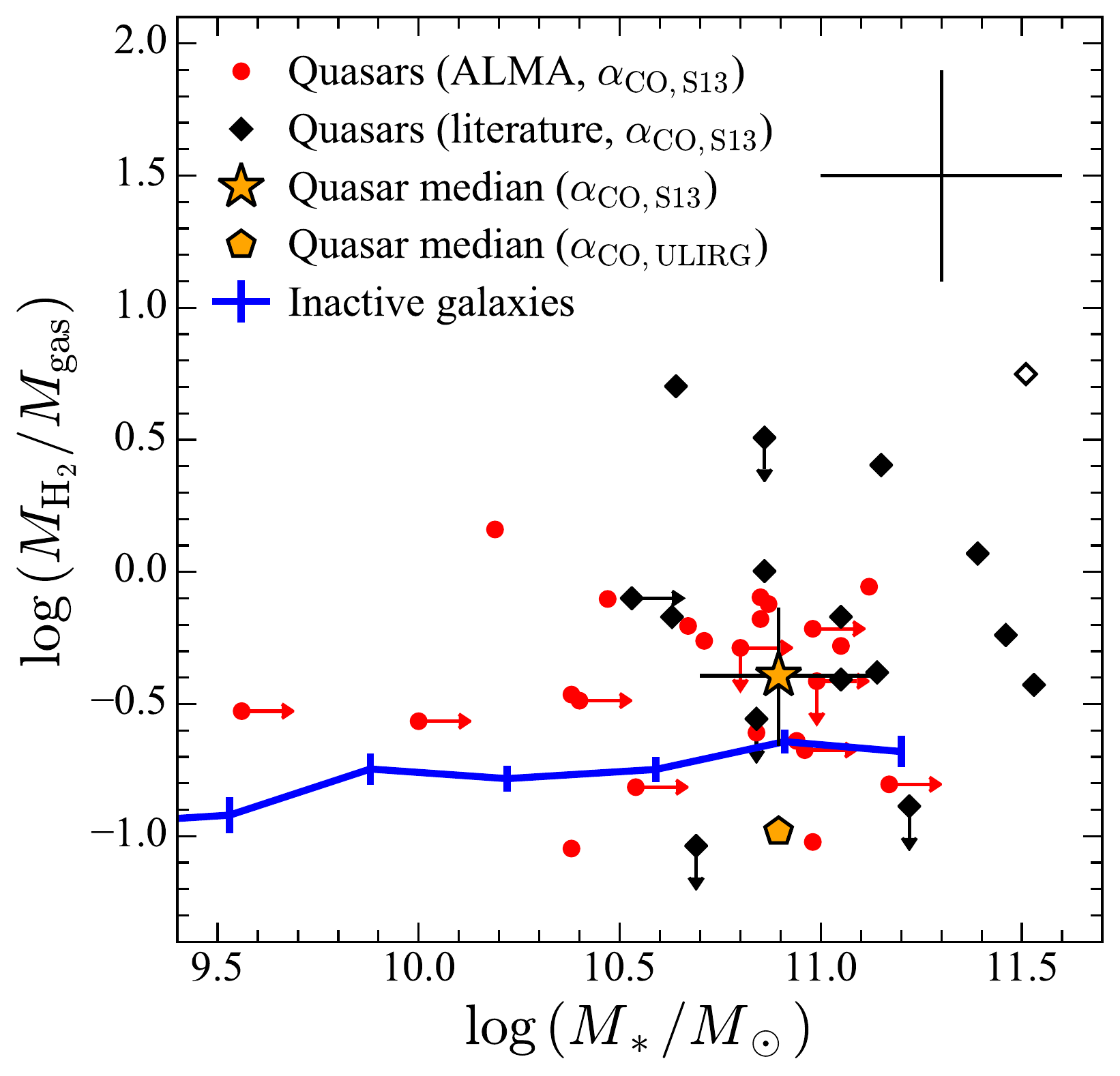}
\caption{
The molecular-to-total gas mass ratios of quasars are consistent with those 
of inactive galaxies, within the scatter.  The total gas mass is estimated from 
the dust mass \citep{Shangguan2018ApJ}.  The relation between 
$M_\mathrm{H_2}/M_\mathrm{gas}$ and stellar mass for inactive galaxies 
(blue line) is derived from \cite{Catinella2018MNRAS}.  The median and 
$\pm25$th percentiles of $M_\mathrm{H_2}/M_\mathrm{gas}$ and $M_*$ for 
the quasars, accounting for the censored data, are shown as the orange star, 
where we have assumed the \comol\ conversion factor of star-forming galaxies 
from \citet[][$\alpha_{\rm CO,S13}=3.1\,\uaco$]{Sandstrom2013ApJ}.  The median 
gas mass ratio calculated assuming a conversion factor appropriate for ULIRGs 
(orange pentagon; $\alpha_{\rm CO,ULIRG}=0.8\,\uaco$) is significantly lower 
than that for inactive galaxies.  We exclude PG~1226+023 (open diamond) 
in calculating the median values, as its total gas mass is very uncertain.  
The uncertainties of the x-axis ($\sim 0.3$ dex) and y-axis ($\sim 0.4$ dex) are 
indicated in the upper-right corner.
}
\label{fig:aco} 
\end{center}
\end{figure}

Figure \ref{fig:aco} plots the variation of the molecular gas fraction 
($M_{\rm H_2}/M_{\rm gas}$) as a function of stellar mass ($M_\star$), where 
$M_{\rm gas}$ is the total mass of the cold interstellar medium ($M_{\rm H~I} + 
M_{\rm H_2}$) inferred from the dust mass, as described in 
\cite{Shangguan2018ApJ}.  The host galaxies of PG quasars,\footnote{For 
the purposes of this discussion, we exclude PG~1545+210, whose molecular 
gas mass and total gas mass are upper limits.} accounting for the censored 
data, have a $50\pm25$ percentile molecular gas fraction of $40\% \pm 24\%$ 
and a stellar mass of $10^{10.89\pm0.22}\,M_\odot$.\footnote{Again, 
PG~1226+023 is excluded because its total gas mass, derived from 
the dust mass, is very uncertain (see Table \ref{tab:lite} and 
\citealt{Shangguan2018ApJ}).  However, the median values are barely affected by
this choice.}  
The quasars gathered from the literature on average have a higher molecular 
gas fraction than those newly observed using ALMA.  This is an obvious 
observational selection effect.  If we limit ourselves to the unbiased ALMA sample, 
the $50\pm25$ percentile molecular gas fraction becomes $32\% \pm 18\%$ 
for a stellar mass of $10^{10.86 \pm 0.19}\,M_\odot$.  The molecular gas fraction 
of the quasars are in rough agreement with, but slightly more elevated than, 
that of inactive galaxies of similar stellar mass (\citealt{Catinella2018MNRAS}; 
blue line in Figure \ref{fig:aco}).\footnote{\cite{Catinella2018MNRAS} 
used a variable \aco, which is on average $\sim3.0$ \uaco\ for galaxies with 
$M_*>10^{10.5}\,M_\odot$, very close to our value.}  This is not unexpected.  
AGNs in general and quasars in particular reside preferentially in 
bulge-dominated galaxies \citep{Ho1997ApJ,Ho2008ARAA,Kim2017ApJS,Zhao2019ApJ}, 
and bulge-dominated systems tend to have higher molecular gas fractions 
\citep{Catinella2018MNRAS}.  In other words, at any given stellar mass, AGN 
hosts, by virtue of their earlier type morphologies, should have higher 
molecular gas fractions, as observed.

The above analysis, while far from a rigorous derivation, does suggest that 
the host galaxies of low-redshift quasars have an \aco\ value not too dissimilar 
from that of ordinary star-forming galaxies and lower luminosity AGNs.  We do 
not believe that the \comol\ conversion factor of PG quasars can be as low as 
\aco\ = 0.8 \uaco, a value commonly advocated for ultraluminous IR galaxies 
(ULIRGs; \citealt{Downes1998ApJ}).  Such a low value of \aco\ would result in 
molecular gas mass fractions substantially lower than those of star-forming 
galaxies (orange pentagon in  Figure \ref{fig:aco}).  This seems improbable.
As shown in Section \ref{ssec:lcoir}, quasar host galaxies follow nearly the
same \lir--\lco\ relation as starburst galaxies, suggesting that they have 
similarly high star formation efficiencies.

\section{Summary}
\label{sec:sum}

We present new ALMA Compact Array observations of the CO(2--1) line for 
23 $z<0.1$ Palomar-Green quasars.  We detect CO(2--1) emission in 21 
objects---13 for the first time---and provide stringent upper limits for the 
remaining two, almost doubling the number of PG quasars with CO detections.  
Combined with published CO(1--0) observations, we assemble CO 
measurements for a representative sample of 40 $z<0.3$ PG quasars, 
which forms the basis of a companion investigation on the relations 
between AGN properties and the molecular gas properties of quasar host 
galaxies \citep{Shangguan2019bApJ}.

This work, primarily devoted to the observational aspects of the new ALMA 
observations and the general characteristics of the sample, highlights the 
following results:  

\begin{itemize}

\item The CO(2--1)/CO(1--0) ratio of low-redshift quasar host galaxies,
$R_{21}=0.62^{+0.15}_{-0.07}$, is broadly consistent with that 
of low-redshift star-forming and active galaxies.  The molecular gas is likely 
subthermal.  We do not find a strong correlation between $R_{21}$ and the mean
intensity of the interstellar radiation field or AGN luminosity.
 
\item Quasar host galaxies follow a tight, linear \lir--\lco\ relation that 
strongly resembles the behavior of starburst galaxies.  
 
\item 
Quasar host galaxies have molecular-to-total gas mass fractions slightly higher 
than, but generally consistent with, those of normal galaxies, if the
\comol\ conversion factor is that of nearby star-forming galaxies, 
$\alpha_\mathrm{CO}=3.1\,M_\odot\,\mathrm{(K\,km\,s^{-1}\,pc^{2})^{-1}}$.

\end{itemize}

\acknowledgments
We are grateful to an anonymous referee for helpful comments and 
suggestions.  We acknowledge support from: the National Science Foundation 
of China grant 11721303 (LCH), the National Key R\&D Program of China grant 
2016YFA0400702 (LCH); CONICYT-Chile grants Basal AFB-170002 (FEB, ET), 
FONDO ALMA 31160033 (FEB), FONDECYT Regular 1160999 (ET) and 1190818 
(ET, FEB), and Anillo de ciencia y tecnologia ACT1720033 (ET); and the Chilean 
Ministry of Economy, Development, and Tourism’s Millennium Science Initiative 
through grant IC120009, awarded to The Millennium Institute of Astrophysics, 
MAS (FEB).  JS thanks Feng Long, Jiayi Sun, Ming-Yang Zhuang, Yali Shao, and 
Jianan Li for helpful discussions.  He is also grateful to Yulin Zhao for sharing the 
\galfit\ results of the PG quasar host galaxies.  Hassen Yusef provided valuable 
advice on statistical methods.  This paper makes use of the following ALMA data: 
ADS/JAO.ALMA\#2017.1.00297.S. ALMA is a partnership of ESO (representing 
its member states), NSF (USA) and NINS (Japan), together with NRC (Canada), 
MOST and ASIAA (Taiwan), and KASI (Republic of Korea), in cooperation with 
the Republic of Chile. The Joint ALMA Observatory is operated by ESO, AUI/NRAO 
and NAOJ.

\vspace{5mm}
\facilities{ALMA}

\software{CASA \citep{McMullin2007ASP}, astropy \citep{Astropy2013AA}, 
PyRAF\footnote{PyRAF is a product of the Space Telescope Science Institute, 
which is operated by AURA for NASA.}
}

\appendix
\section{CO(2--1) Measurements for Individual Objects}
\label{apd:plt}

We detected CO(2--1) emission in 21 out of 23 PG quasars.  Figure \ref{fig:data1} 
shows the moment 0 maps and one-dimensional spectra for 20 objects; the data for 
PG~0050+124 appear in Figure \ref{fig:drex}.  The signal-to-noise ratio of PG~1341+258 
is too low to robustly fit its line profile.  We believe PG~1341+258 is marginally
detected, because we always detect the source with $\sim 4\,\sigma$ significance when 
we clean the data with different velocity channel widths.

\begin{figure}
\begin{center}
\includegraphics[width=0.48\textwidth]{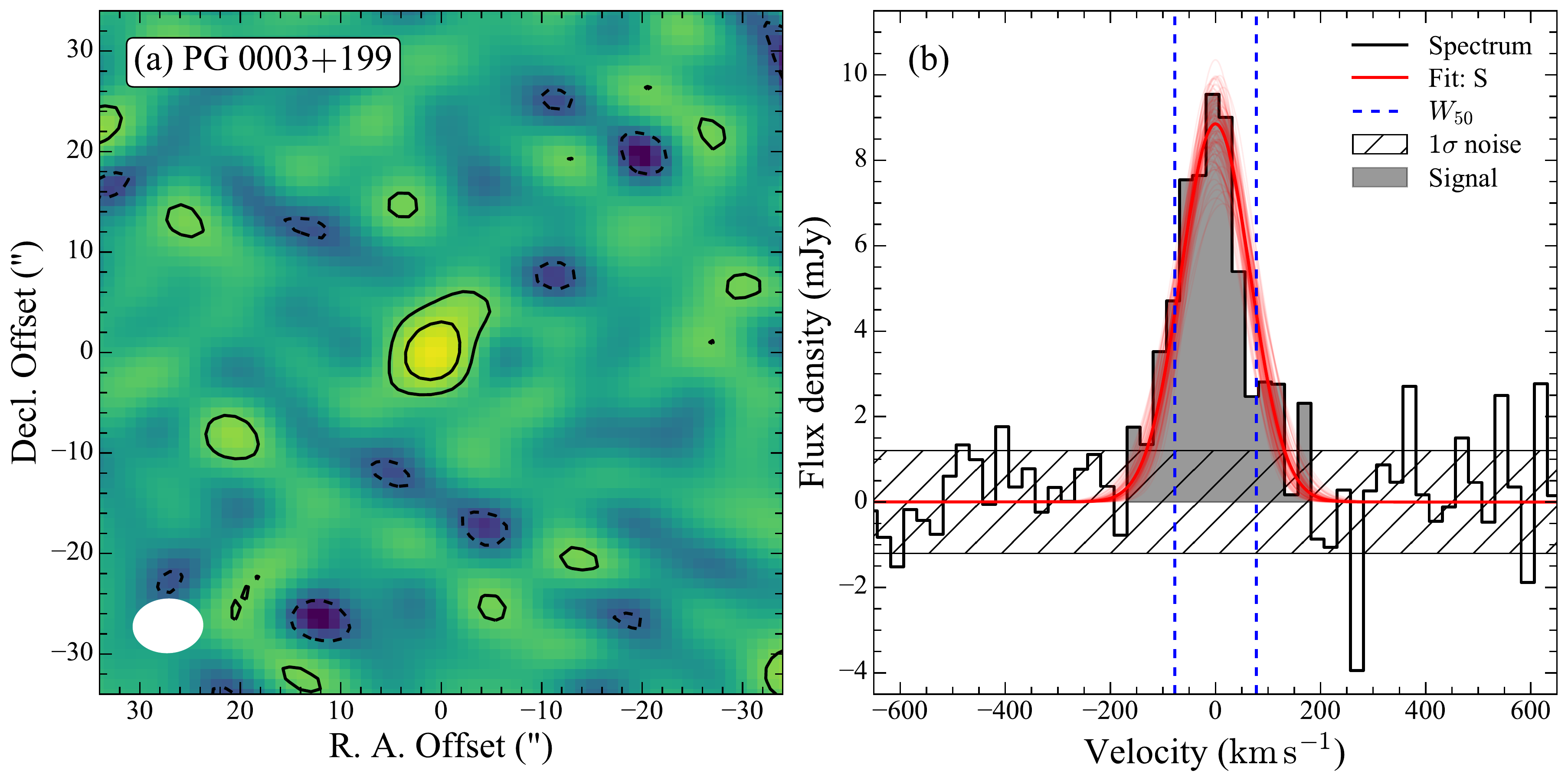}
\includegraphics[width=0.48\textwidth]{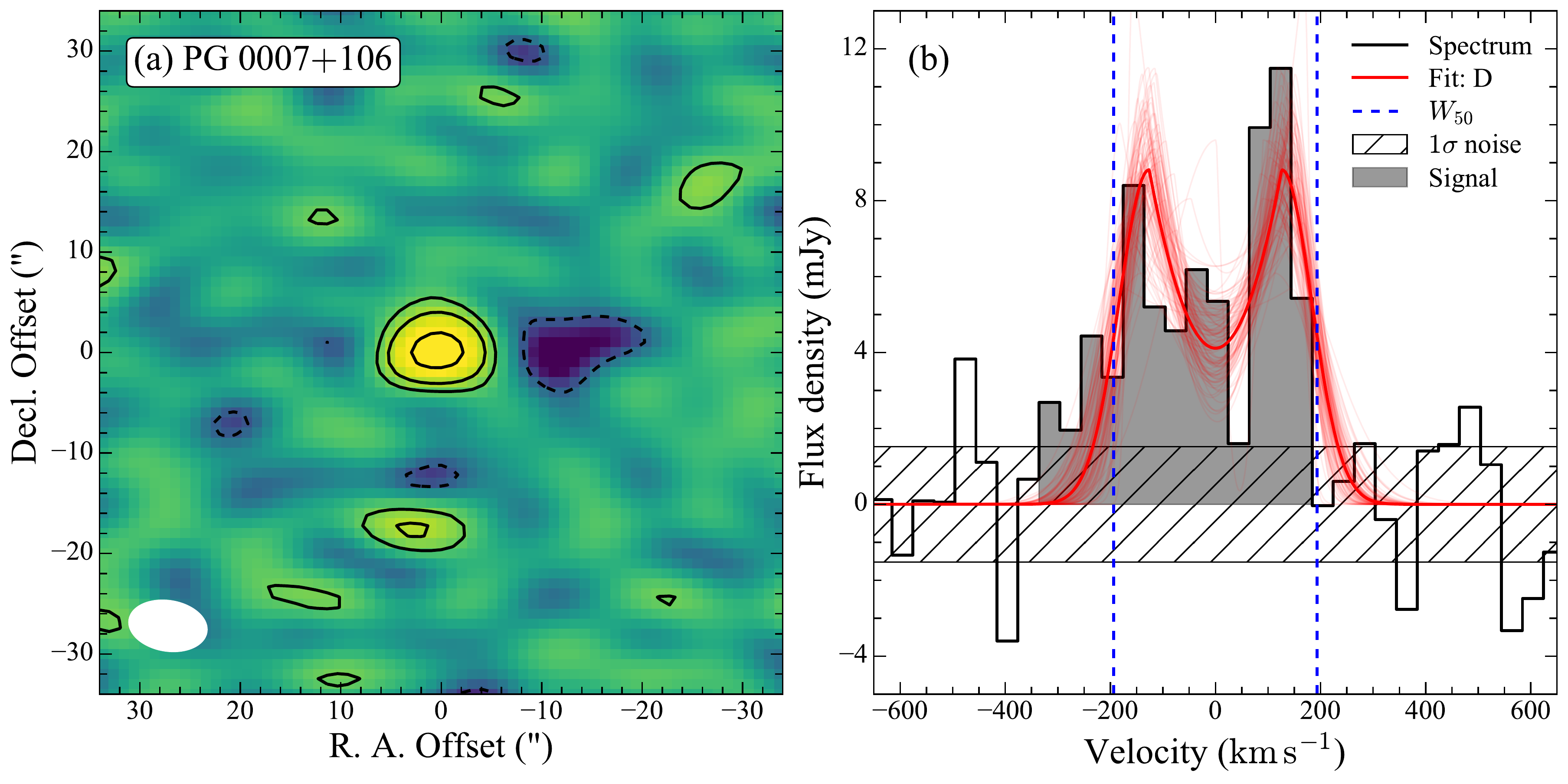}
\includegraphics[width=0.48\textwidth]{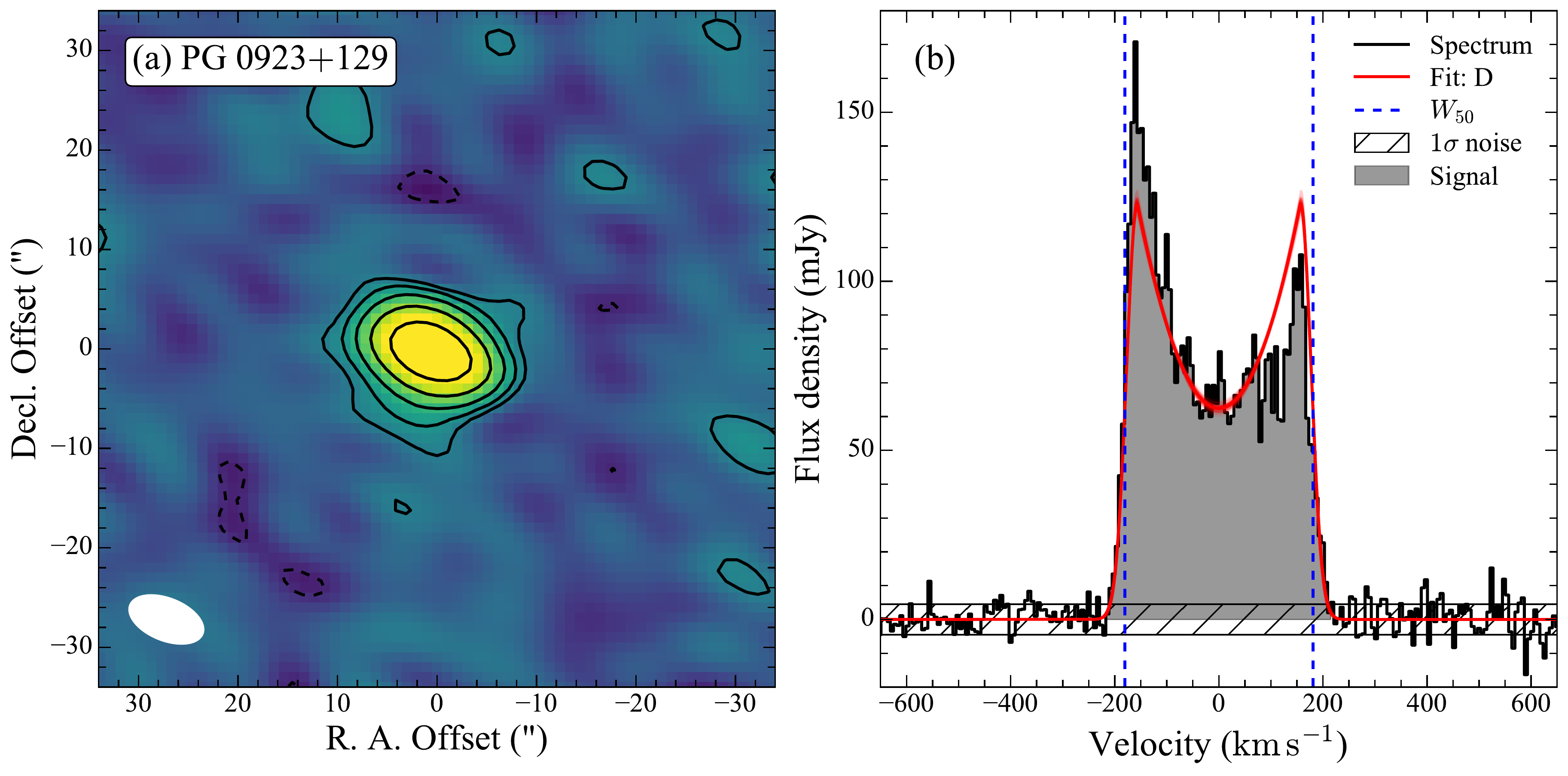}
\includegraphics[width=0.48\textwidth]{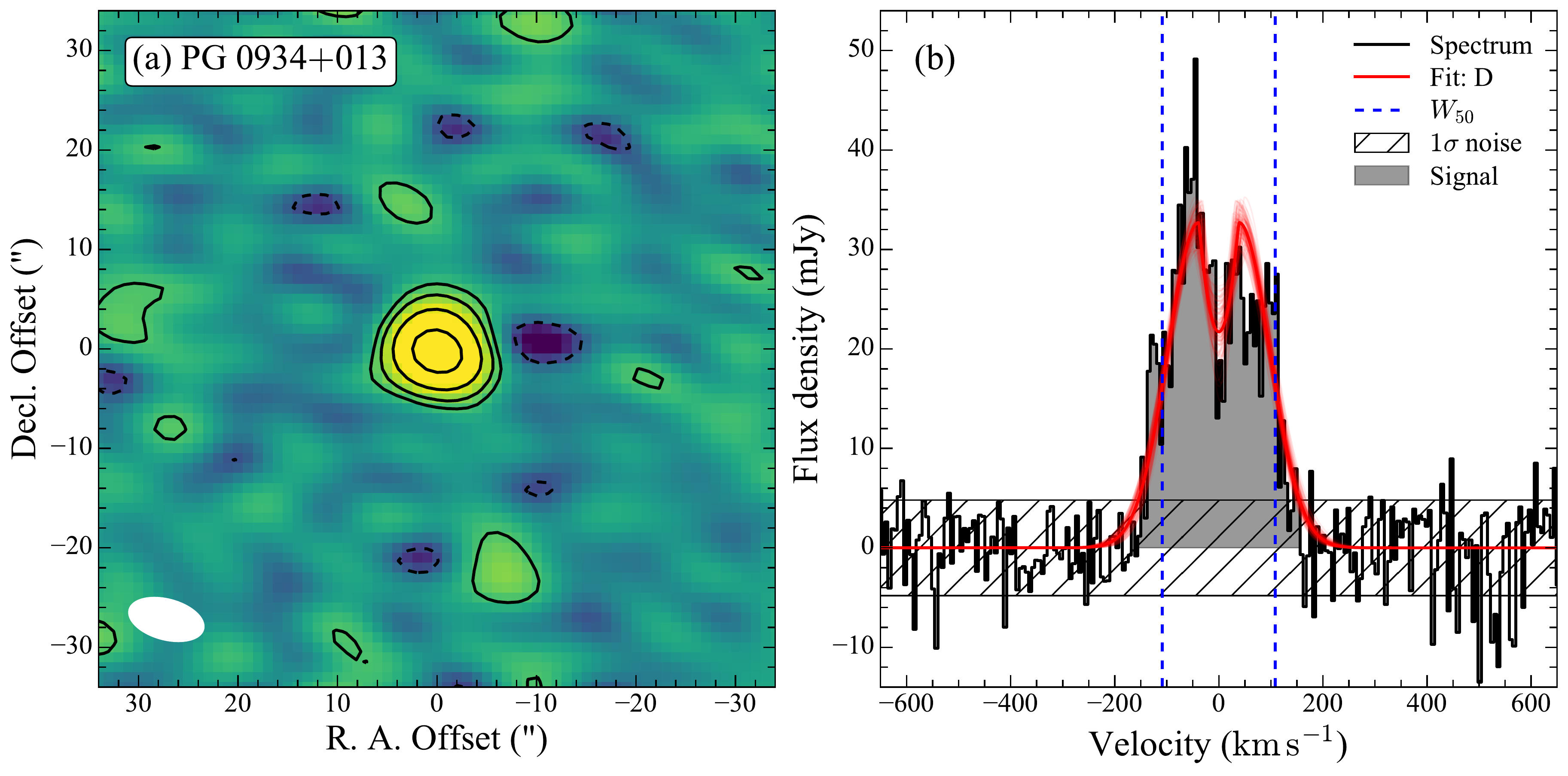}
\includegraphics[width=0.48\textwidth]{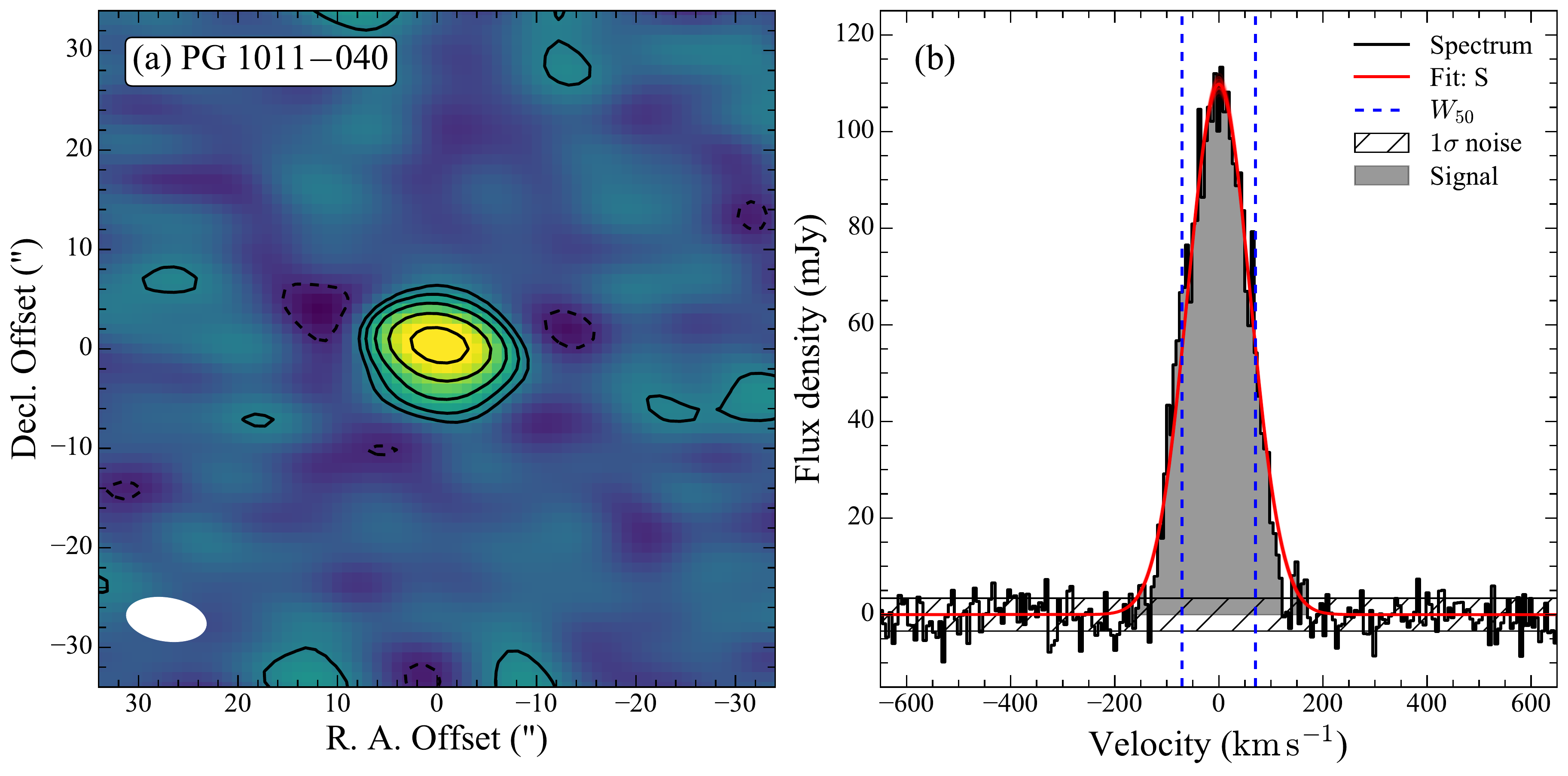}
\includegraphics[width=0.48\textwidth]{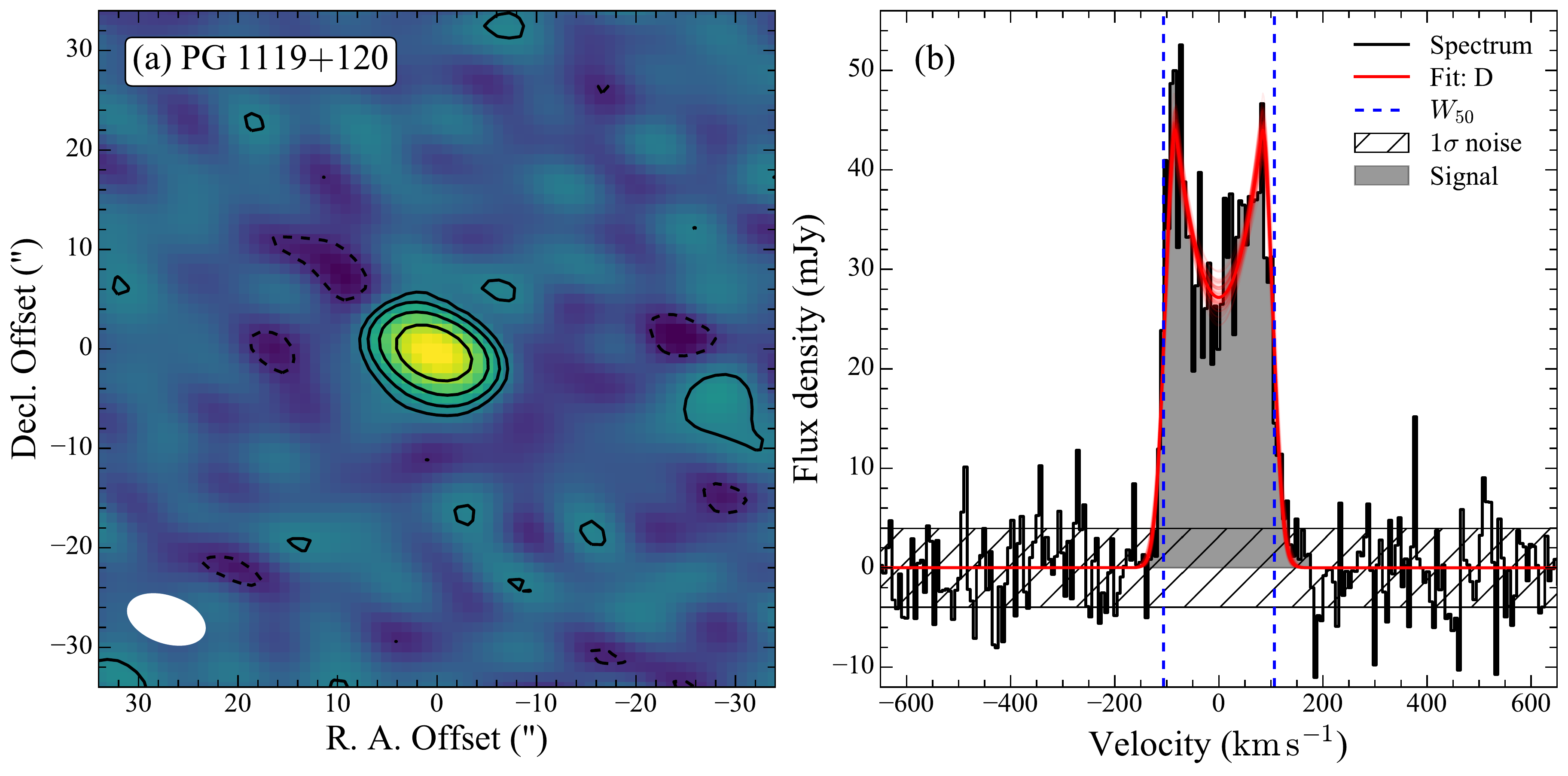}
\includegraphics[width=0.48\textwidth]{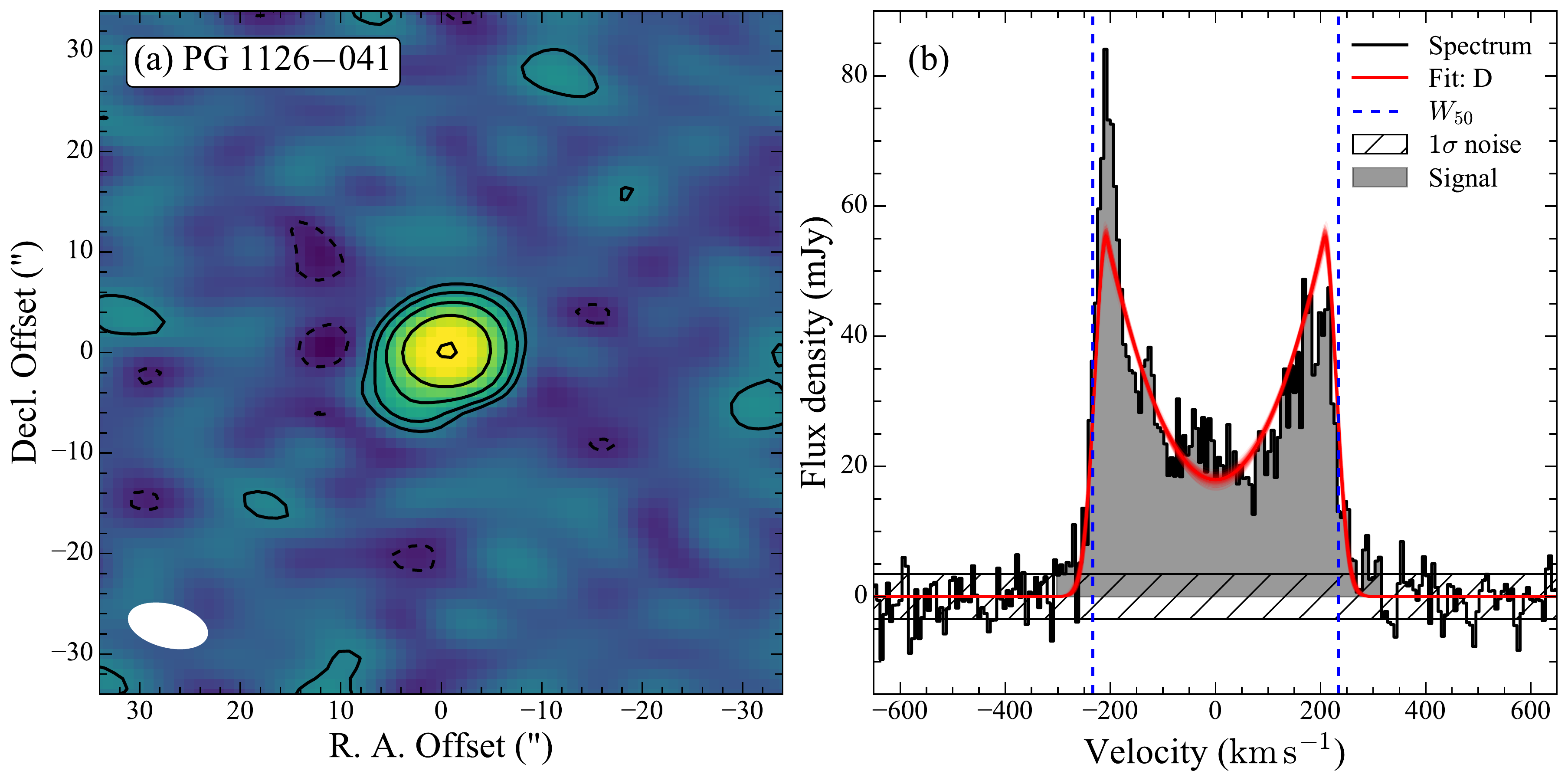}
\includegraphics[width=0.48\textwidth]{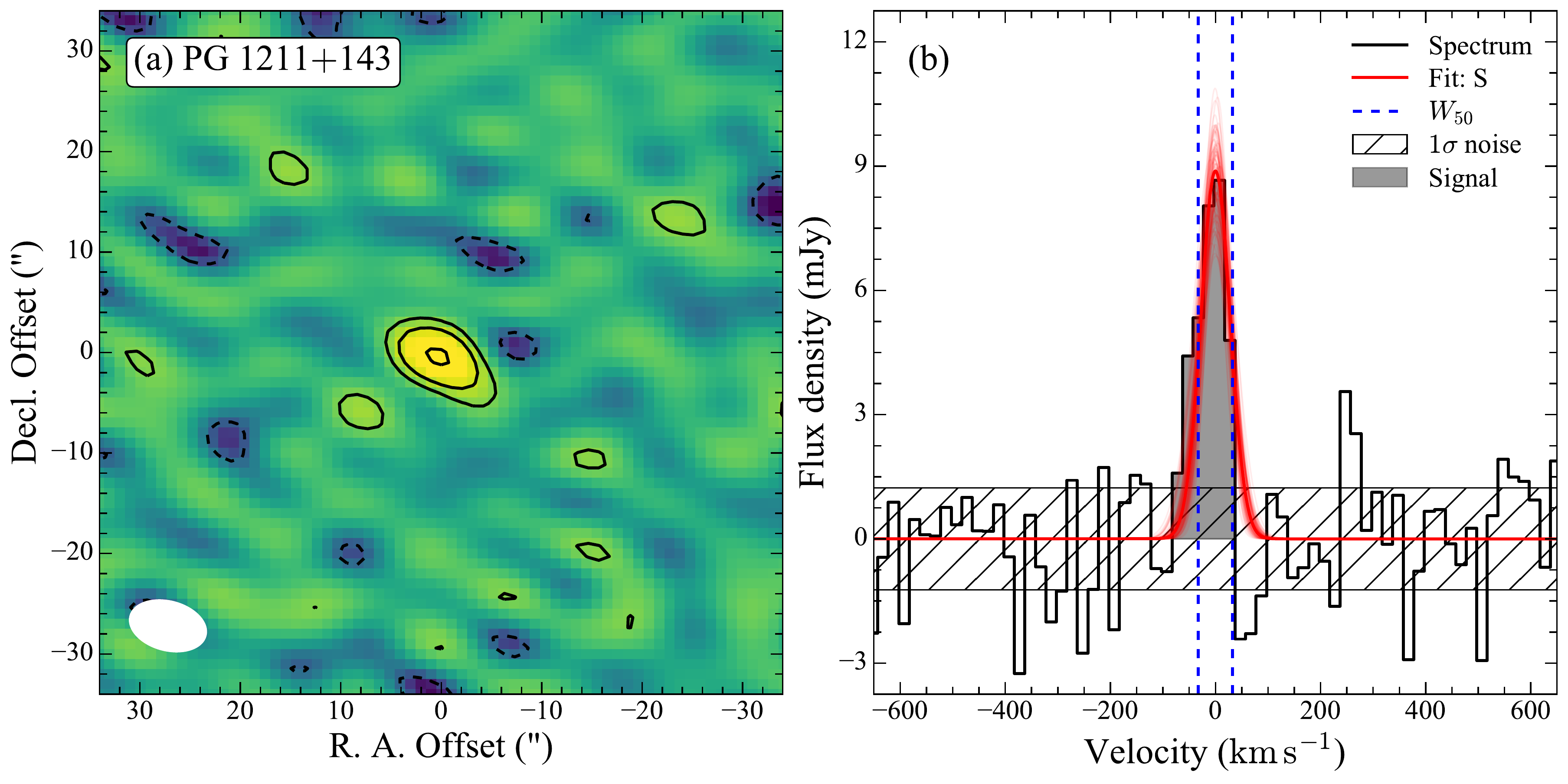}
\includegraphics[width=0.48\textwidth]{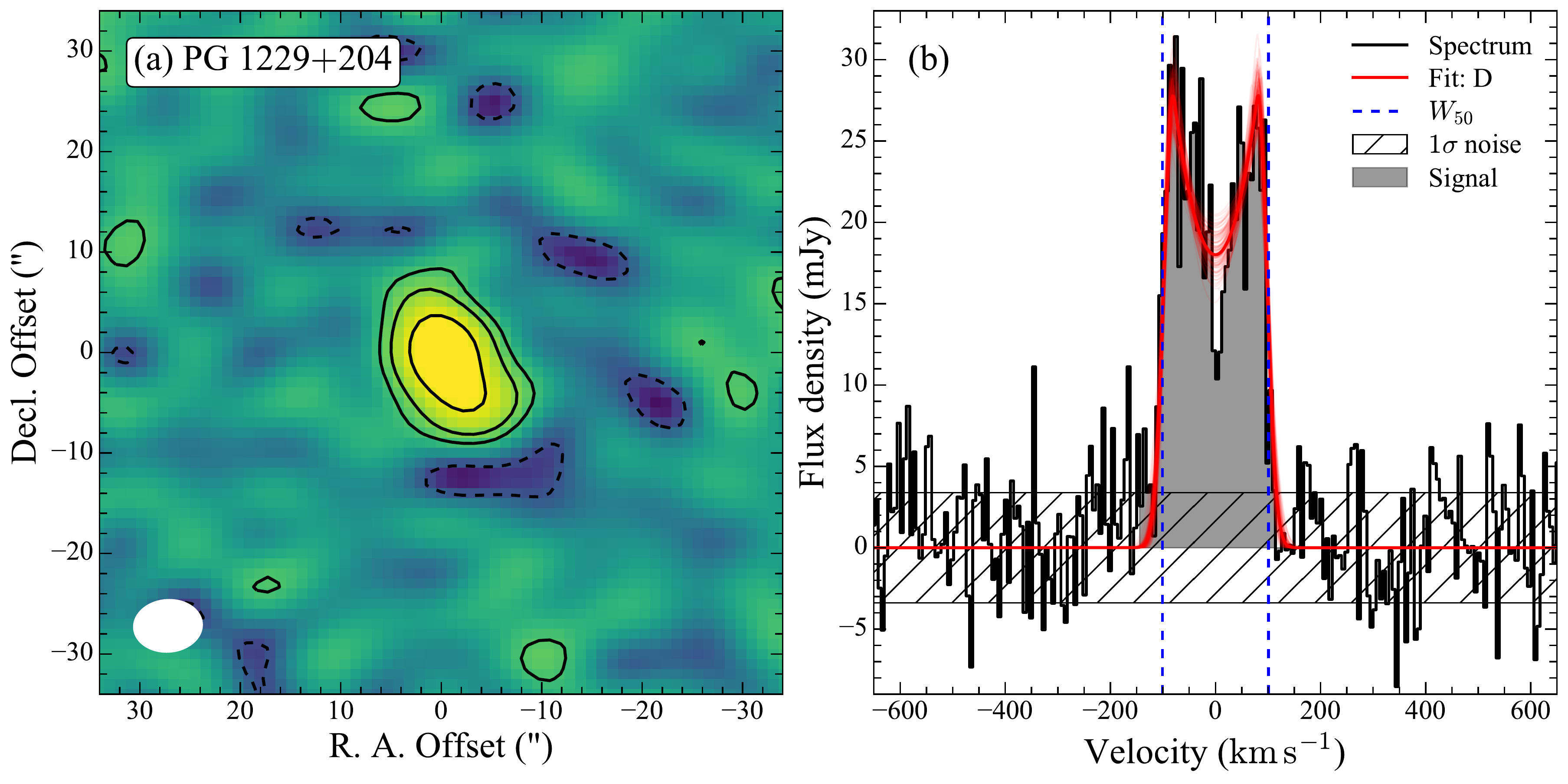}
\includegraphics[width=0.48\textwidth]{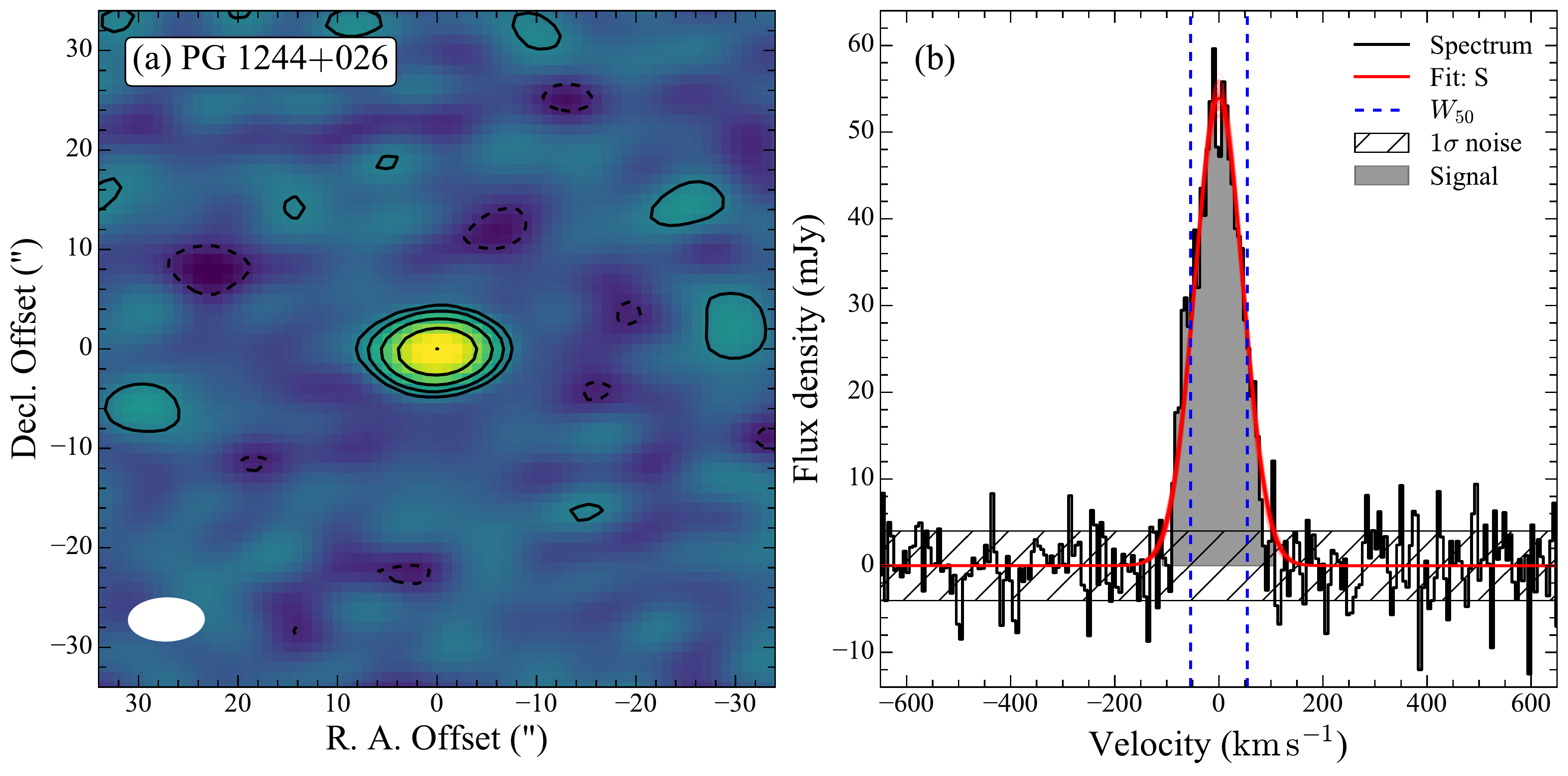}
\caption{Objects in the survey detected in CO(2--1).  (a) Intensity (moment~0) map.  The contours 
are $-$2 (dashed), 2, 4, 8, 16, and 32 $\sigma$ levels, with $\sigma$ being the rms of the source-free 
pixels in the map.  The synthesis beam is indicated on the lower-left corner of the map.  (b) The 
one-dimensional spectrum extracted from the 2 $\sigma$ contour of the source emission.  Channels 
shaded in grey are considered to be signal from the emission line.  The hatched horizontal band 
indicates the noise level of the line-free channels.  The best-fit emission-line profile is plotted 
with a red curve, with the uncertainty displayed with faint thin red lines.  The double-peaked Gaussian 
profile is indicated with ``Fit: D'', while the single Gaussian profile is indicated with ``Fit: S''.  The 
full width of the 50 percentile of the best-fit profile, $W_{50}$, is indicated by the blue dashed lines.}
\label{fig:data1}
\end{center}
\end{figure}

\begin{figure}
\figurenum{6}
\begin{center}
\includegraphics[width=0.48\textwidth]{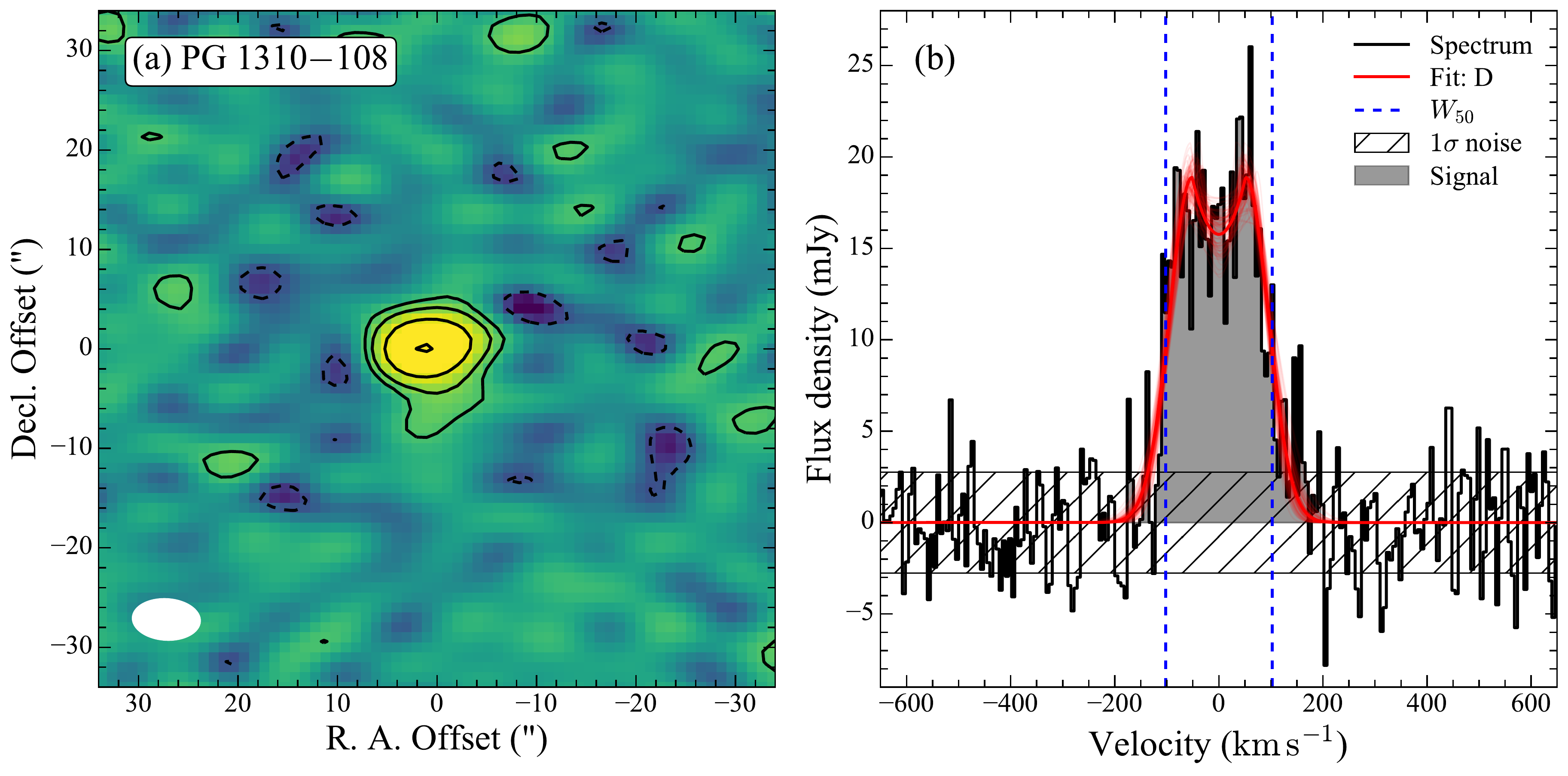}
\includegraphics[width=0.48\textwidth]{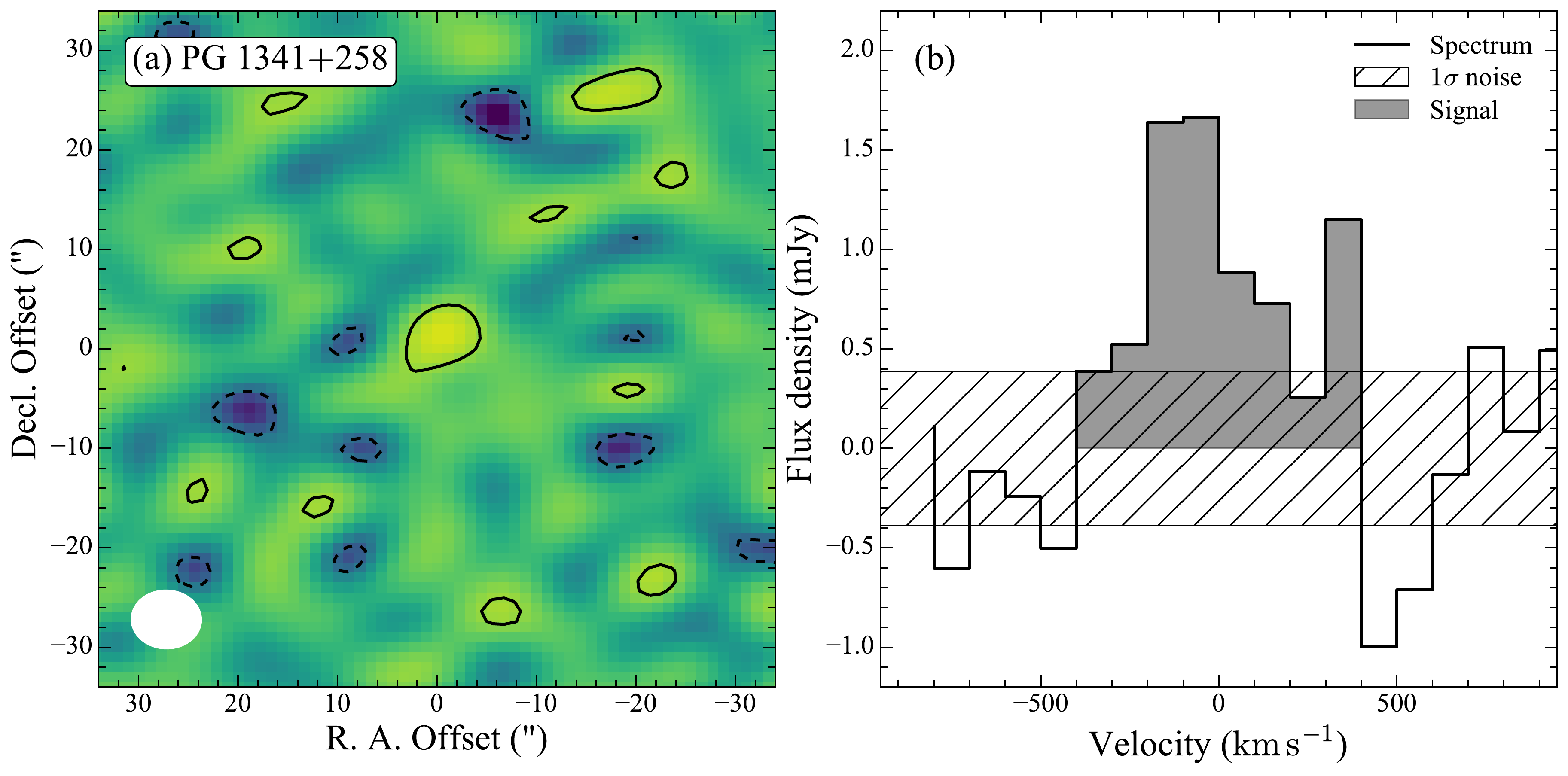}
\includegraphics[width=0.48\textwidth]{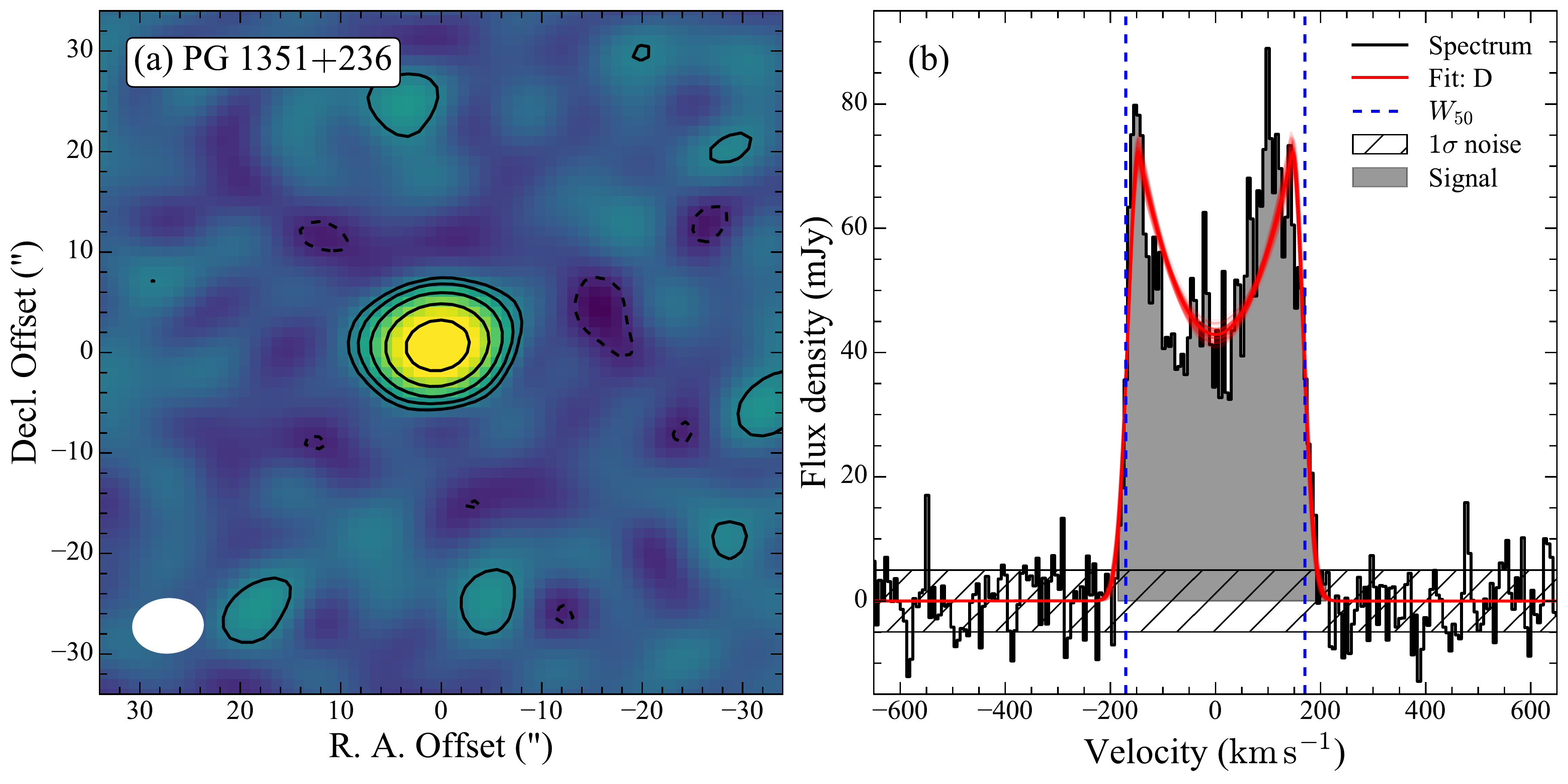}
\includegraphics[width=0.48\textwidth]{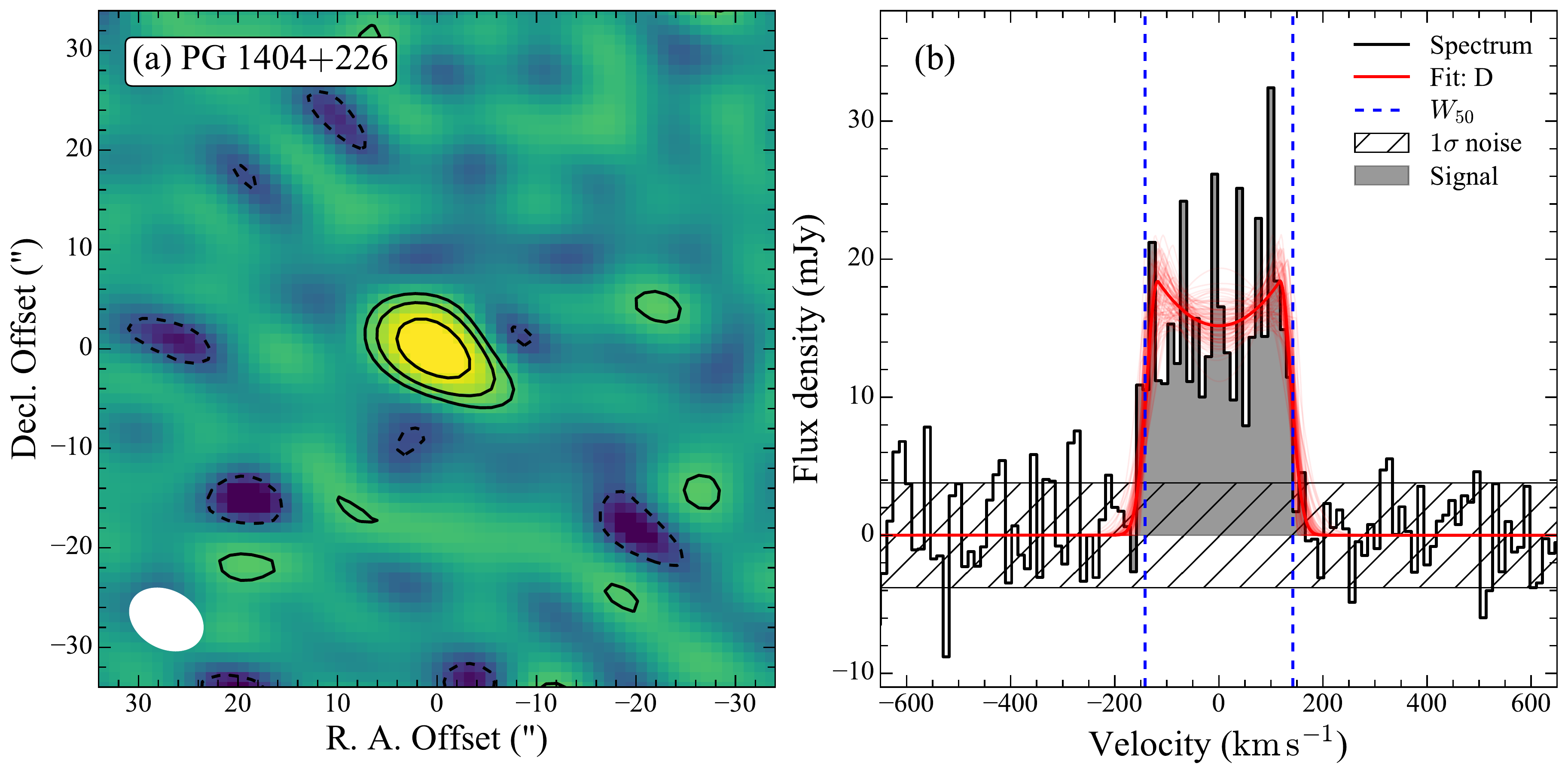}
\includegraphics[width=0.48\textwidth]{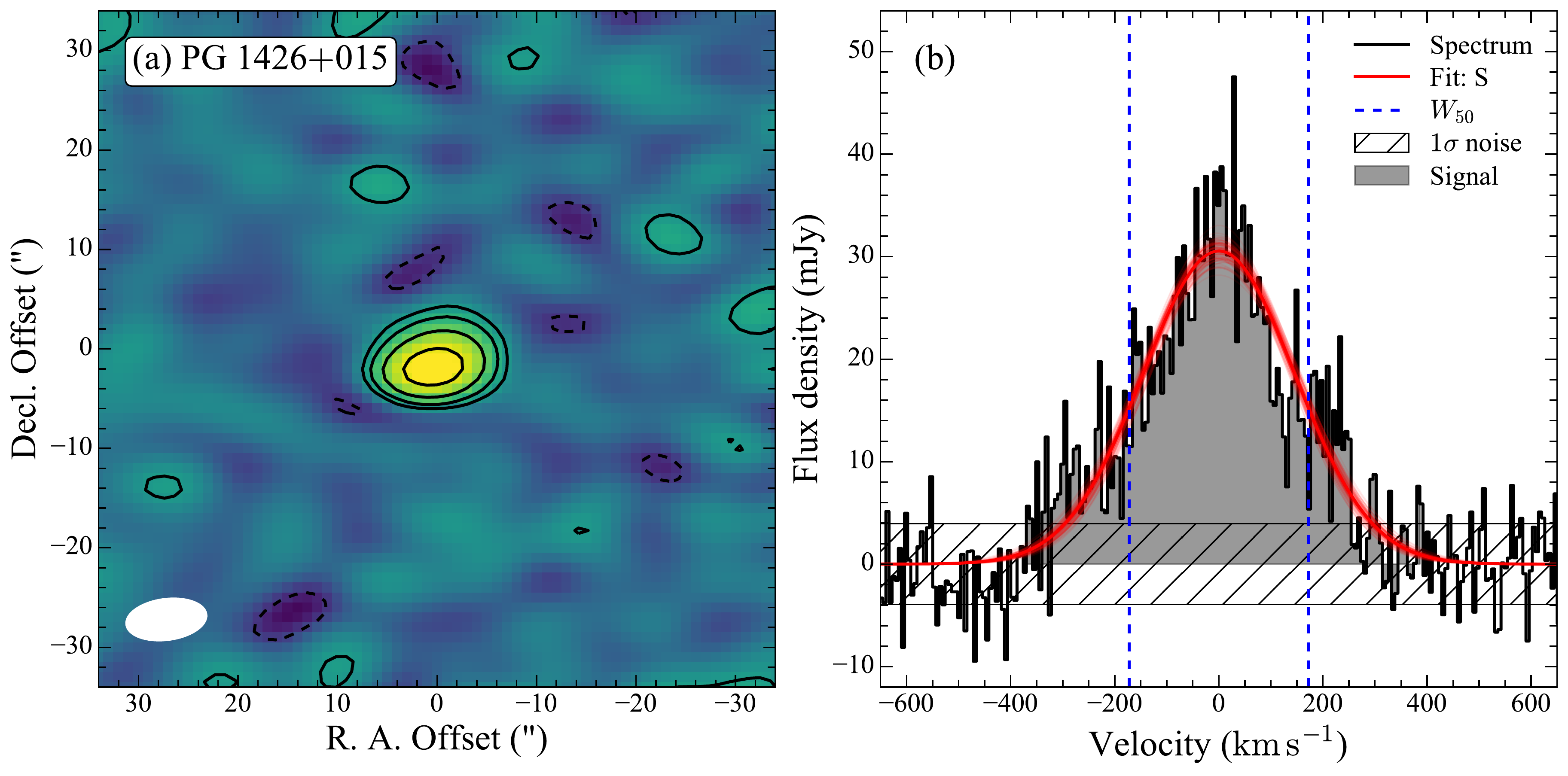}
\includegraphics[width=0.48\textwidth]{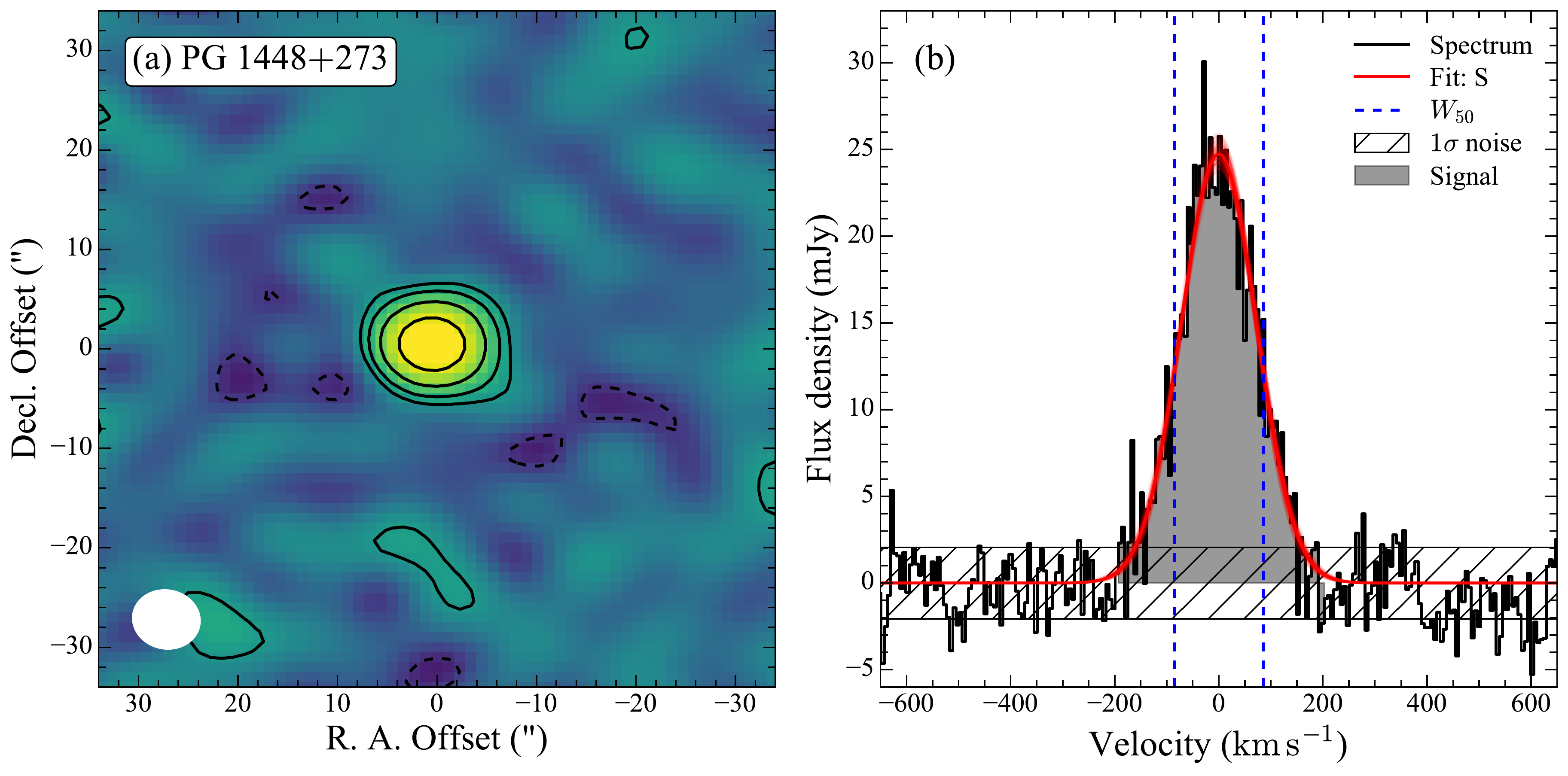}
\includegraphics[width=0.48\textwidth]{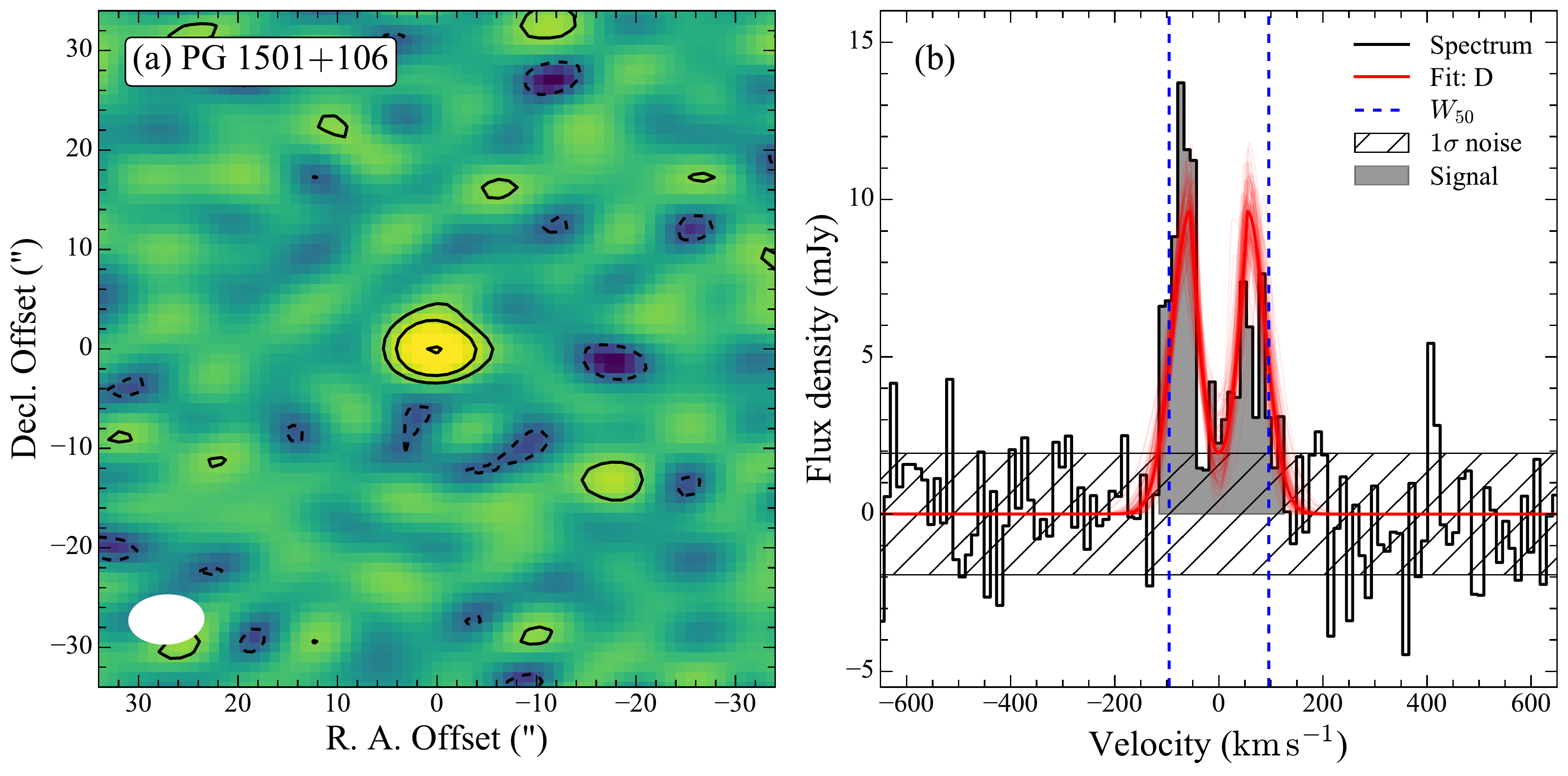}
\includegraphics[width=0.48\textwidth]{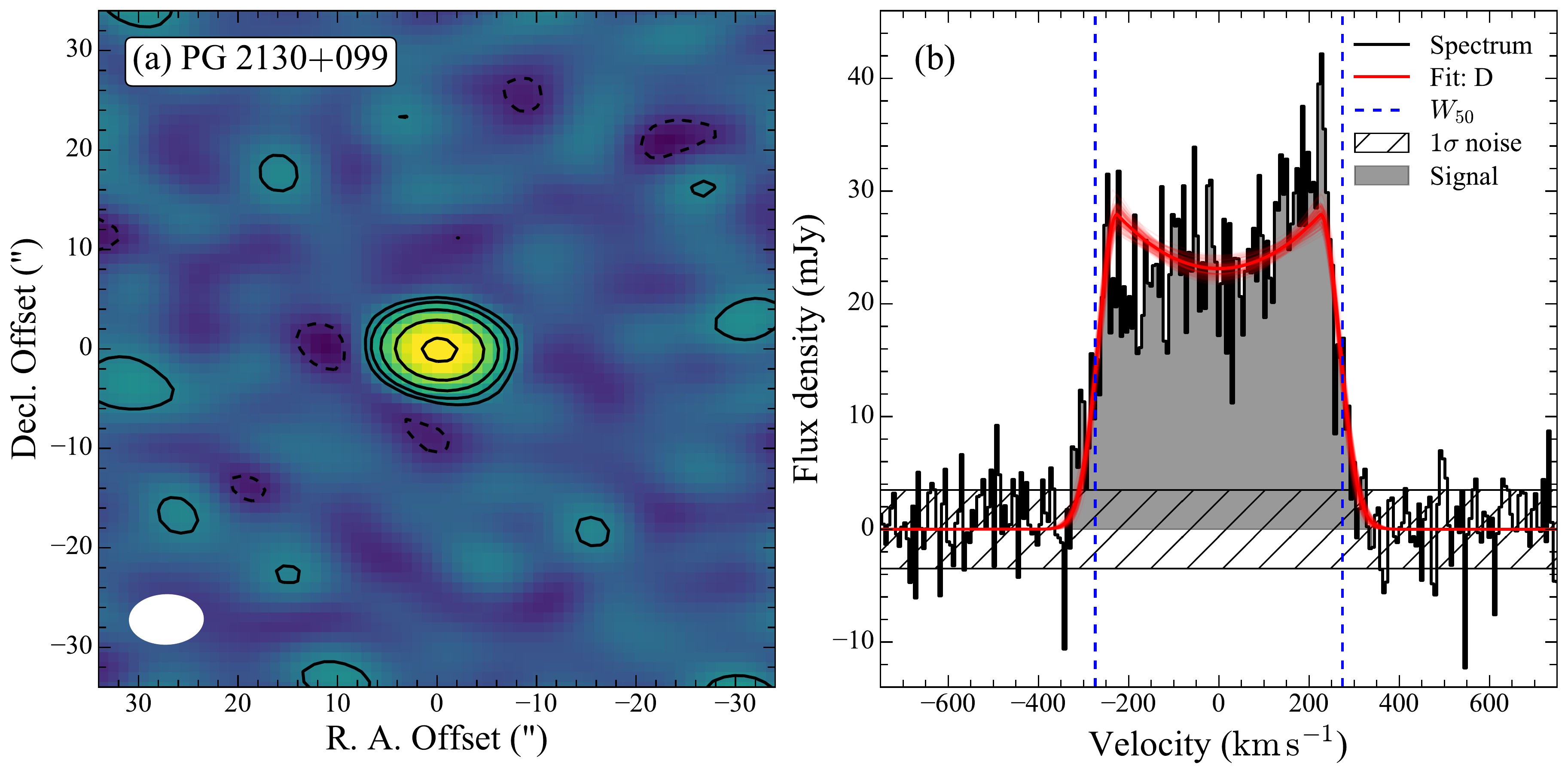}
\includegraphics[width=0.48\textwidth]{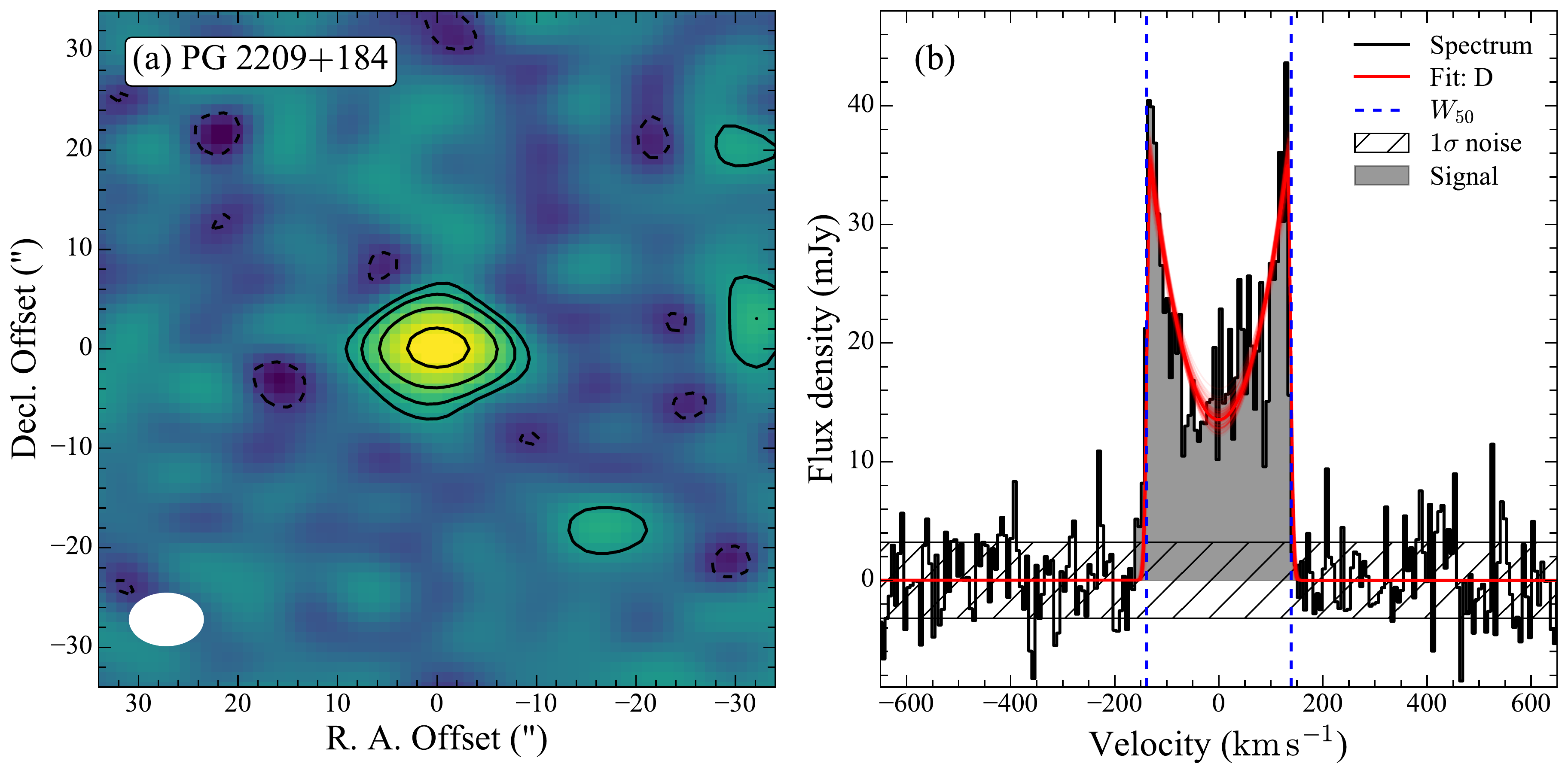}
\includegraphics[width=0.48\textwidth]{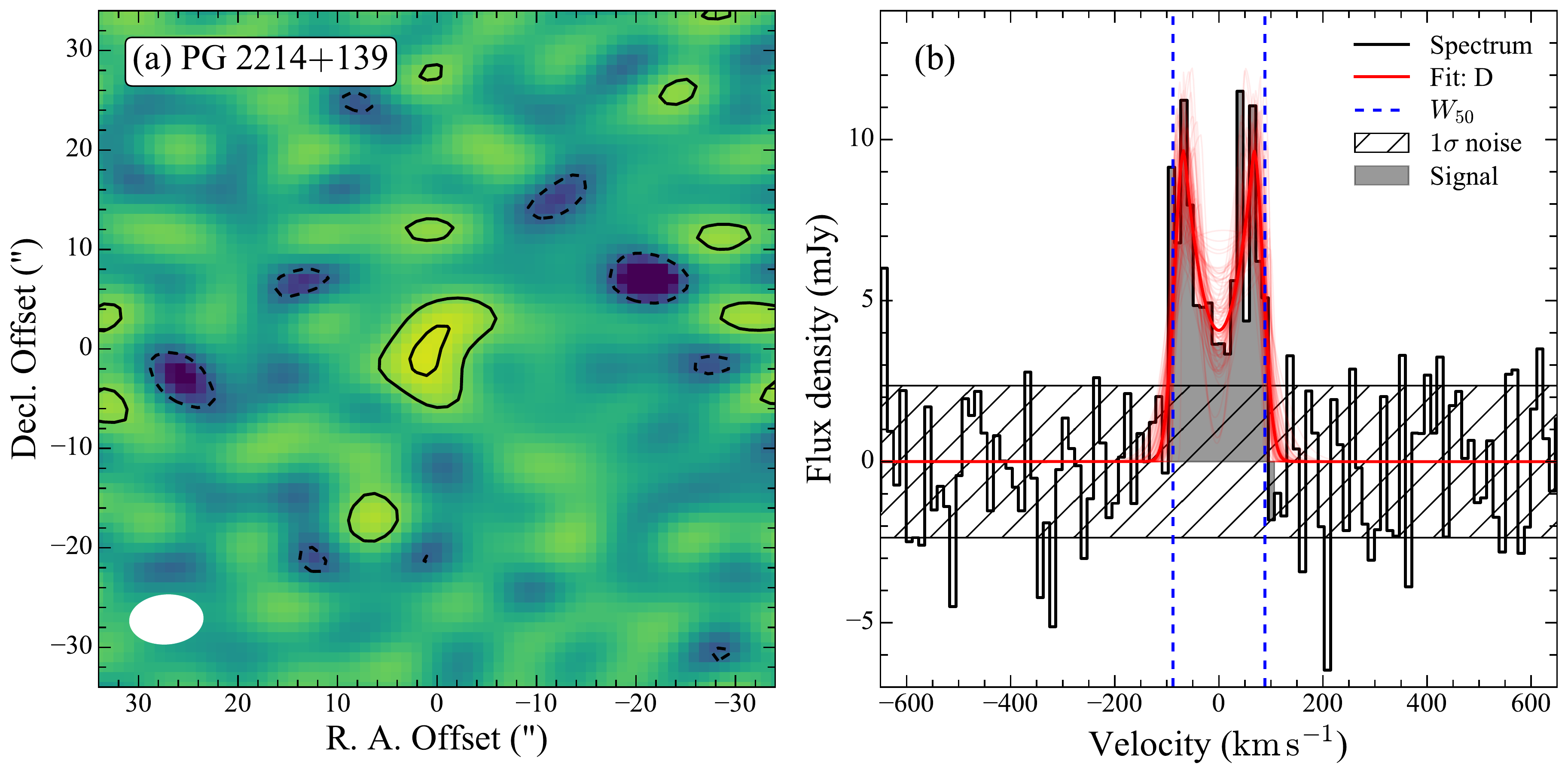}
\caption{Continued.}
\end{center}
\end{figure}

\begin{deluxetable}{c c c c l l c c}
\tablecaption{PG  Quasars ALMA Observations \label{tab:obs}}
\tabletypesize{\scriptsize}
\tablehead{
\colhead{Object} &
\colhead{$z$} &
\colhead{R.A.} &
\colhead{Decl.} &
\colhead{Flux \& Bandpass} &
\colhead{Phase} &
\colhead{PWV} &
\colhead{$T_\mathrm{int}$}\\
\colhead{} &
\colhead{} &
\colhead{(J2000.0)} &
\colhead{(J2000.0)} &
\colhead{Calibrator} &
\colhead{Calibrator} &
\colhead{(mm)} &
\colhead{(minute)}
}
\colnumbers
\startdata
PG 0003$+$199 & 0.025 & 00:06:19.52 & $+$20:12:10.5 &              J2253$+$1608 &              J0019$+$2021 & $0.769\pm0.166$ & 150.58 \\
PG 0007$+$106 & 0.089 & 00:10:31.01 & $+$10:58:29.5 &              J2253$+$1608 & J0010$+$1724/J0022$+$0608 & $2.314\pm0.108$ & 150.69 \\
PG 0049$+$171 & 0.064 & 00:51:54.80 & $+$17:25:58.4 &              J2253$+$1608 &              J0019$+$2021 & $1.222\pm0.100$ & 152.27 \\
PG 0050$+$124 & 0.061 & 00:53:34.94 & $+$12:41:36.2 &              J2253$+$1608 &              J0121$+$1149 & $1.079\pm0.118$ & 152.16 \\
PG 0923$+$129 & 0.029 & 09:26:03.29 & $+$12:44:03.6 & J0522$-$3627/J1058$+$0133 &              J0854$+$2006 & $1.353\pm0.011$ & 152.22 \\
PG 0934$+$013 & 0.050 & 09:37:01.03 & $+$01:05:43.5 & J1058$+$0133/J0854$+$2006 &              J0948$+$0022 & $0.506\pm0.020$ & 149.05 \\
PG 1011$-$040 & 0.058 & 10:14:20.69 & $-$04:18:40.5 &              J1058$+$0133 & J1010$-$0200/J0942$-$0759 & $0.656\pm0.004$ & 151.13 \\
PG 1119$+$120 & 0.049 & 11:21:47.10 & $+$11:44:18.3 &              J1058$+$0133 &              J1116$+$0829 & $1.233\pm0.024$ & 149.00 \\
PG 1126$-$041 & 0.060 & 11:29:16.66 & $-$04:24:07.6 &              J1058$+$0133 &              J1131$-$0500 & $0.720\pm0.069$ & 150.63 \\
PG 1211$+$143 & 0.085 & 12:14:17.70 & $+$14:03:12.6 &              J1229$+$0203 &              J1215$+$1654 & $2.268\pm0.067$ & 150.65 \\
PG 1229$+$204 & 0.064 & 12:32:03.60 & $+$20:09:29.2 &              J1229$+$0203 &              J1224$+$2122 & $3.089\pm0.112$ & 149.25 \\
PG 1244$+$026 & 0.048 & 12:46:35.25 & $+$02:22:08.8 & J1256$-$0547/J1058$+$0133 &              J1229$+$0203 & $0.839\pm0.018$ & 148.98 \\
PG 1310$-$108 & 0.035 & 13:13:05.78 & $-$11:07:42.4 & J1256$-$0547/J1337$-$1257 & J1337$-$1257/J1256$-$0547 & $0.471\pm0.092$ & 150.81 \\
PG 1341$+$258 & 0.087 & 13:43:56.75 & $+$25:38:47.7 &              J1229$+$0203 &              J1333$+$2725 & $1.482\pm0.054$ & 150.63 \\
PG 1351$+$236 & 0.055 & 13:54:06.43 & $+$23:25:49.1 &              J1229$+$0203 &              J1357$+$1919 & $0.974\pm0.055$ & 121.01 \\
PG 1404$+$226 & 0.098 & 14:06:21.89 & $+$22:23:46.6 &              J1229$+$0203 &              J1357$+$1919 & $0.900\pm0.002$ & 151.37 \\
PG 1426$+$015 & 0.086 & 14:29:06.59 & $+$01:17:06.5 & J1337$-$1257/J1256$-$0547 & J1408$-$0752/J1410$+$0203 & $2.158\pm0.010$ & 151.22 \\
PG 1448$+$273 & 0.065 & 14:51:08.76 & $+$27:09:26.9 & J1337$-$1257/J1229$+$0203 & J1446$+$1721/J1427$+$2348 & $0.642\pm0.025$ & 279.44 \\
PG 1501$+$106 & 0.036 & 15:04:01.20 & $+$10:26:16.2 & J1517$-$2422/J1229$+$0203 &              J1504$+$1029 & $2.153\pm0.051$ & 150.69 \\
PG 2130$+$099 & 0.061 & 21:32:27.81 & $+$10:08:19.5 &              J2253$+$1608 &              J2147$+$0929 & $1.498\pm0.075$ & 124.17 \\
PG 2209$+$184 & 0.070 & 22:11:53.89 & $+$18:41:49.9 &              J2253$+$1608 &              J2232$+$1143 & $1.541\pm0.162$ & 152.28 \\
PG 2214$+$139 & 0.067 & 22:17:12.26 & $+$14:14:20.9 &              J2253$+$1608 &              J2232$+$1143 & $1.471\pm0.071$ & 150.71 \\
PG 2304$+$042 & 0.042 & 23:07:02.91 & $+$04:32:57.2 &              J2253$+$1608 & J2327$+$0940/J2320$+$0513 & $0.435\pm0.011$ & 151.23 \\
\enddata
\tablecomments{
(1) Source name.
(2) Redshift.
(3) Right ascension.
(4) Declination.
(5) Bandpass and flux calibrators;  since each source is observed multiple
times, more than one calibrator may be used for one source.
(6) Phase calibrators.
(7) The median and standard deviation of the precipitable water vapour (PWV).
(8) Total on-source integration time.
}
\end{deluxetable}

\begin{longrotatetable}
\begin{deluxetable}{c c c r c c r@{$\pm$}l r c r@{$\pm$}l l l c c r@{$\pm$}l}
\tablecaption{Summary of Observational Results\label{tab:alma}}
\tabletypesize{\tiny}
\tablehead{
\colhead{Object} &
\colhead{$\log\,\lambda L_\lambda$(5100 \AA)} &
\colhead{$\log\,M_\mathrm{BH}$} &
\colhead{$\log\,M_*$} &
\colhead{$q$} &
\colhead{Ref.} &
\mcl{2}{c}{$S_\mathrm{CO}\Delta \nu$} &
\colhead{$\Delta\,S_\mathrm{CO}\Delta \nu$} &
\colhead{$\log\,L^\prime_\mathrm{CO\mbox{\tiny (1--0)}}$} &
\mcl{2}{c}{$\log\,M_\mathrm{H2}$} &
\colhead{$W_{50}$} &
\colhead{$W_{20}$} &
\colhead{Prof.} &
\colhead{$\log\,L_\mathrm{IR}$} &
\mcl{2}{c}{$\log\,M_\mathrm{gas}$}\\
\colhead{} &
\colhead{$(\mathrm{erg\,s^{-1}})$} &
\colhead{$(M_\odot)$} &
\colhead{$(M_\odot)$} &
\colhead{} &
\colhead{} &
\mcl{2}{c}{$(\mathrm{Jy\,km\,s^{-1}})$} &
\colhead{($\sigma_\mathrm{total}$)} &
\colhead{$(\mathrm{K\,km\,s^{-1}\,pc^{2}})$} &
\mcl{2}{c}{$(M_\odot)$} &
\colhead{$(\mathrm{km\,s^{-1}})$} &
\colhead{$(\mathrm{km\,s^{-1}})$} &
\colhead{} &
\colhead{$\mathrm{(erg\,s^{-1})}$} &
\mcl{2}{c}{$(M_\odot)$} \\
\colhead{(1)} &
\colhead{(2)} &
\colhead{(3)} &
\colhead{(4)} &
\colhead{(5)} &
\colhead{(6)} &
\mcl{2}{c}{(7)} &
\colhead{(8)} &
\colhead{(9)} &
\mcl{2}{c}{(10)} &
\colhead{(11)} &
\colhead{(12)} &
\colhead{(13)} &
\colhead{(14)} &
\mcl{2}{c}{(15)}
}
\startdata
PG 0003$+$199 & 44.17 & 7.52 & $>$10.00 &    0.93 &       1 &        1.49 & 0.19  & $-0.058$ &        7.26$\pm$0.06 &         7.75 & 0.30 & $155.06^{+16.37}_{-14.67}$      & $237.51^{+22.54}_{-21.95}$ & S       & $43.11^{+0.03}_{-0.03}$ &  8.32 & 0.20 \\
PG 0007$+$106 & 44.79 & 8.87 &    10.84 & \nodata & \nodata &        2.85 & 0.31  & $-1.052$ &        8.66$\pm$0.05 &         9.15 & 0.30 & $386.78^{+29.57}_{-25.18}$      & $460.52^{+47.60}_{-42.04}$ & D       & $44.27^{+0.02}_{-0.03}$ &  9.76 & 0.22 \\
PG 0049$+$171 & 43.97 & 8.45 & $>$10.80 & \nodata & \nodata & \mcl{2}{c}{$<$0.88} & \nodata  &              $<$7.86 & \mcl{2}{c}{$<$8.35} & \mcl{1}{c}{\nodata}             & \mcl{1}{c}{\nodata}        & \nodata & $42.91^{+0.05}_{-0.08}$ &  8.66 & 0.36 \\
PG 0050$+$124 & 44.76 & 7.57 &    11.12 &    0.53 &       2 &       75.99 & 0.80  & $-0.827$ &        9.75$\pm$0.01 &        10.24 & 0.30 & $377.77^{+0.85}_{-0.86}$        & $431.59^{+1.47}_{-1.32}$   & D       & $44.94^{+0.01}_{-0.01}$ & 10.30 & 0.20 \\
PG 0923$+$129 & 43.83 & 7.52 &    10.71 &    0.78 &       2 &       32.24 & 0.87  & $ 2.113$ &        8.73$\pm$0.01 &         9.22 & 0.30 & $361.68^{+1.07}_{-1.03}$        & $387.27^{+1.87}_{-1.71}$   & D       & $44.05^{+0.01}_{-0.02}$ &  9.48 & 0.20 \\
PG 0934$+$013 & 43.85 & 7.15 &    10.38 &    0.69 &       2 &        6.72 & 0.39  & $ 0.218$ &        8.52$\pm$0.03 &         9.02 & 0.30 & $217.84^{+7.98}_{-7.15}$        & $290.73^{+10.24}_{-10.58}$ & D       & $43.96^{+0.02}_{-0.02}$ &  9.48 & 0.20 \\
PG 1011$-$040 & 44.23 & 7.43 &    10.87 &    0.92 &       2 &       16.20 & 0.44  & $-0.205$ &        9.04$\pm$0.01 &         9.53 & 0.30 & $141.00^{+1.42}_{-1.35}$        & $214.90^{+2.09}_{-2.04}$   & S       & $43.98^{+0.02}_{-0.02}$ &  9.65 & 0.20 \\
PG 1119$+$120 & 44.10 & 7.58 &    10.67 &    0.63 &       2 &        7.67 & 0.28  & $-1.050$ &        8.56$\pm$0.02 &         9.06 & 0.30 & $212.68^{+2.41}_{-2.37}$        & $236.69^{+3.78}_{-3.87}$   & D       & $44.12^{+0.02}_{-0.04}$ &  9.26 & 0.20 \\
PG 1126$-$041 & 44.36 & 7.87 &    10.85 & \nodata & \nodata &       16.04 & 0.64  & $-0.592$ &        9.06$\pm$0.02 &         9.55 & 0.30 & $467.00^{+1.74}_{-1.72}$        & $494.33^{+3.21}_{-3.25}$   & D       & $44.46^{+0.03}_{-0.03}$ &  9.65 & 0.20 \\
PG 1211$+$143 & 45.04 & 8.10 &    10.38 &    0.84 &       1 &        0.64 & 0.05  & $-0.666$ &        7.97$\pm$0.03 &         8.46 & 0.30 & $\ph{0}65.90^{+7.29}_{-7.00}$   & $100.94^{+11.41}_{-11.49}$ & S       & $43.32^{+0.05}_{-0.05}$ &  9.51 & 0.29 \\
PG 1229$+$204 & 44.35 & 8.26 &    10.94 &    0.55 &       1 &        4.73 & 0.32  & $ 0.894$ &        8.59$\pm$0.03 &         9.08 & 0.30 & $202.21^{+3.14}_{-2.84}$        & $223.50^{+5.55}_{-4.53}$   & D       & $43.96^{+0.01}_{-0.01}$ &  9.72 & 0.20 \\
PG 1244$+$026 & 43.77 & 6.62 &    10.19 &    0.70 &       2 &        6.14 & 0.26  & $-2.003$ &        8.45$\pm$0.02 &         8.94 & 0.30 & $108.94^{+2.94}_{-2.91}$        & $166.18^{+4.33}_{-4.75}$   & S       & $43.85^{+0.02}_{-0.01}$ &  8.78 & 0.20 \\
PG 1310$-$108 & 43.70 & 7.99 & $>$10.40 & \nodata & \nodata &        3.87 & 0.15  & $ 0.064$ &        7.97$\pm$0.02 &         8.46 & 0.30 & $204.08^{+7.04}_{-6.33}$        & $258.09^{+10.67}_{-9.77}$  & D       & $43.16^{+0.02}_{-0.01}$ &  8.95 & 0.20 \\
PG 1341$+$258 & 44.31 & 8.15 & $>$10.54 & \nodata & \nodata &        0.67 & 0.15  & $ 0.173$ &        8.01$\pm$0.10 &         8.51 & 0.31 & \mcl{1}{c}{\nodata}             & \mcl{1}{c}{\nodata}        & \nodata & $43.81^{+0.04}_{-0.05}$ &  9.32 & 0.25 \\
PG 1351$+$236 & 44.02 & 8.67 & $>$10.98 & \nodata & \nodata &       19.17 & 0.67  & $-2.093$ &        9.06$\pm$0.02 &         9.55 & 0.30 & $340.88^{+1.84}_{-1.81}$        & $366.46^{+2.77}_{-3.03}$   & D       & $44.28^{+0.01}_{-0.01}$ &  9.77 & 0.20 \\
PG 1404$+$226 & 44.35 & 7.01 &  $>$9.56 & \nodata & \nodata &        4.80 & 0.26  & $-0.584$ &        8.97$\pm$0.02 &         9.46 & 0.30 & $284.69^{+9.27}_{-8.31}$        & $312.29^{+12.40}_{-11.86}$ & D       & $43.97^{+0.02}_{-0.02}$ &  9.99 & 0.20 \\
PG 1426$+$015 & 44.85 & 9.15 &    11.05 & \nodata &       1 &       11.31 & 0.56  & $-1.343$ &        9.23$\pm$0.02 &         9.72 & 0.30 & $343.71^{+10.26}_{-10.16}$      & $524.34^{+14.77}_{-17.54}$ & S       & $44.55^{+0.02}_{-0.02}$ & 10.00 & 0.20 \\
PG 1448$+$273 & 44.45 & 7.09 &    10.47 &    0.63 &       2 &        4.45 & 0.21  & $-0.374$ &        8.58$\pm$0.02 &         9.07 & 0.30 & $170.78^{+4.07}_{-4.27}$        & $260.13^{+6.28}_{-6.91}$   & S       & $43.95^{+0.02}_{-0.02}$ &  9.17 & 0.20 \\
PG 1501$+$106 & 44.26 & 8.64 & $>$10.96 & \nodata & \nodata &        1.30 & 0.12  & $-1.416$ &        7.52$\pm$0.04 &         8.02 & 0.30 & $192.68^{+10.41}_{-9.49}$       & $234.10^{+16.94}_{-14.66}$ & D       & $43.74^{+0.07}_{-0.05}$ &  8.69 & 0.20 \\
PG 2130$+$099 & 44.54 & 8.04 &    10.85 &    0.44 &       1 &       14.07 & 0.39  & $-1.050$ &        9.02$\pm$0.01 &         9.51 & 0.30 & $548.36^{+4.98}_{-4.68}$        & $600.61^{+7.12}_{-8.20}$   & D       & $44.37^{+0.02}_{-0.03}$ &  9.69 & 0.20 \\
PG 2209$+$184 & 44.44 & 8.89 & $>$11.17 & \nodata & \nodata &        6.05 & 0.26  & $ 0.613$ &        8.77$\pm$0.02 &         9.27 & 0.30 & $277.48^{+1.36}_{-1.22}$        & $284.66^{+2.69}_{-2.32}$   & D       & $43.81^{+0.02}_{-0.03}$ & 10.07 & 0.20 \\
PG 2214$+$139 & 44.63 & 8.68 &    10.98 &    0.97 &       2 &        1.24 & 0.13  & $ 2.489$ &        8.05$\pm$0.05 &         8.54 & 0.30 & $179.64^{+9.00}_{-8.89}$        & $200.38^{+17.19}_{-11.13}$ & D       & $43.57^{+0.01}_{-0.02}$ &  9.56 & 0.21 \\
PG 2304$+$042 & 44.04 & 8.68 & $>$10.99 & \nodata & \nodata & \mcl{2}{c}{$<$0.80} & \nodata  &              $<$7.45 & \mcl{2}{c}{$<$7.94} & \mcl{1}{c}{\nodata}             & \mcl{1}{c}{\nodata}        & \nodata & $42.66^{+0.06}_{-0.08}$ &  8.37 & 0.27 \\
\enddata
\tablecomments{
(1) Source name.
(2) AGN monochromatic luminosity of the continuum at 5100 \AA.
(3) BH mass.
(4) Stellar mass of the host galaxy.  The uncertainty of the direct
measurements is $\sim 0.3$ dex.  The lower limits come from bulge masses
estimated from the BH mass using the $M_\mathrm{BH}-M_\mathrm{bulge}$ relation.
See Table 1 of \cite{Shangguan2018ApJ} for more details.
(5) Axial ratio, derived from GALFIT modeling of the quasar host galaxies.
(6) References for the axial ratio.
(7) Integrated line flux of CO(2--1) emission.
(8) The significance of the measured CO(2--1) fluxes deviate from those measured
from the 15\arcsec\ UV-tapered images.  Positive deviation means the tapered flux
larger than the flux in previous column.  $\sigma_\mathrm{total}$ is the quadrature
sum of the flux uncertainties.
(9) CO line luminosity, converted from $J=$ (2--1) to $J=$ (1--0) with a
line ratio of 0.62.
(10) Molecular gas mass derived from CO line luminosity, assuming
$\alpha_\mathrm{CO}=3.1\,M_\odot\,\mathrm{(K\,km\,s^{-1}\,pc^{2})^{-1}}$.
(11) The width of the CO integrated profile at 50 percent of its maximum.
(12) The width of the CO integrated profile at 20 percent of its maximum.
(13) CO line profile: ``S'' = single-peaked profile and ``D'' = double-peaked profile.
(14) IR luminosity of the host galaxy from spectral energy distribution decomposition
by \cite{Shangguan2018ApJ}.
(15) Total gas mass derived from the dust mass.
Columns (2)--(5) and (14) are collected from Table 1 of \cite{Shangguan2018ApJ}.
References: (1) \cite{Kim2017ApJS}; (2) Y. Zhao et al. (in preparation).
}
\end{deluxetable}
\end{longrotatetable}

\begin{table*}
  \begin{center}
  \caption{CO(2--1)/CO(1--0) Line Ratio}
  \label{tab:r21}
  \begin{tabular}{cccccc}
    \hline
    \hline
       Object & $\log\,L^\prime_\mathrm{CO(1-0)}$  & $\log\,L^\prime_\mathrm{CO(2-1)}$  & $R_{21}$      & $\langle U \rangle$ & Ref. \\        %
              & ($\mathrm{K\,km\,s^{-1}\,pc^{2}}$) & ($\mathrm{K\,km\,s^{-1}\,pc^{2}}$) &               &                     &      \\        %
       (1)    & (2)                                & (3)                                & (4)           & (5)                 & (6)  \\ \hline %
PG 0003$+$199 &       $<$8.87                      & 7.05$\pm$0.05                      &       $>$0.02 &         15.16       & 1\tnm{a}\phantom{\hspace{-4.pt}} \\
PG 0007$+$106 &       $<$9.07                      & 8.45$\pm$0.05                      &       $>$0.24 &    \ph{0}7.95       & 2    \\
PG 0050$+$124 & 9.74$\pm$0.02                      & 9.54$\pm$0.01                      & 0.63$\pm$0.02 &         10.07       & 3\tnm{b}\phantom{\hspace{-4.pt}} \\
PG 0934$+$013 &       $<$8.34                      & 8.31$\pm$0.03                      &       $>$0.93 &    \ph{0}7.20       & 4    \\
PG 1011$-$040 & 9.05$\pm$0.04                      & 8.83$\pm$0.01                      & 0.60$\pm$0.05 &    \ph{0}5.10       & 4    \\
PG 1119$+$120 & 8.50$\pm$0.06                      & 8.35$\pm$0.02                      & 0.71$\pm$0.11 &         17.57       & 3    \\
PG 1126$-$041 & 9.12$\pm$0.04                      & 8.85$\pm$0.02                      & 0.53$\pm$0.05 &         15.36       & 4    \\
PG 1211$+$143 &       $<$8.73                      & 7.76$\pm$0.03                      &       $>$0.11 &    \ph{0}2.00       & 5    \\
PG 1229$+$204 & 8.69$\pm$0.11                      & 8.38$\pm$0.03                      & 0.49$\pm$0.13 &    \ph{0}4.02       & 5    \\
PG 1310$-$108 &       $<$8.16                      & 7.76$\pm$0.02                      &       $>$0.40 &    \ph{0}3.98       & 4    \\
PG 1404$+$226 & 8.98$\pm$0.11                      & 8.76$\pm$0.02                      & 0.60$\pm$0.15 &    \ph{0}2.39       & 5    \\
PG 1426$+$015 & 9.12$\pm$0.07                      & 9.02$\pm$0.02                      & 0.79$\pm$0.14 &    \ph{0}8.32       & 5    \\
PG 1501$+$106 &       $<$9.24                      & 7.31$\pm$0.04                      &       $>$0.01 &         25.48       & 1    \\
PG 2130$+$099 & 8.85$\pm$0.06                      & 8.81$\pm$0.01                      & 0.90$\pm$0.12 &         11.04       & 3    \\
PG 2214$+$139 &       $<$8.55                      & 7.84$\pm$0.05                      &       $>$0.19 &    \ph{0}2.55       & 5\tnm{a}\phantom{\hspace{-4.pt}} \\ \hline
  \end{tabular}
  \end{center}
\tabletypesize{\scriptsize}
\tablenotetext{a}{$L^\prime_\mathrm{CO(1-0)}$ is considered an upper
limit; archival measurement has poor S/N.}
\tablenotetext{b}{The line flux and FWHM of PG 0050+124 from \cite{Evans2006AJ} are entirely
consistent with those reported recently by \cite{Tan2019arXiv}.}
\tablecomments{
(1) Object name.
(2) The CO(1--0) line luminosity from the literature.
(3) The CO(2--1) line luminosity from our ALMA observations (see Table \ref{tab:alma}).
(4) The CO line luminosity ratio, $R_{21}\equiv L^\prime_\mathrm{CO(2-1)}/L^\prime_\mathrm{CO(1-0)}$.
(5) The mean interstellar radiation field intensity derived from the IR spectral energy distribution of the quasar \citep{Shangguan2018ApJ}.
(6) References: (1) \cite{Maiolino1997ApJ}; (2) \cite{Evans2001AJ}; (3) \cite{Evans2006AJ};
                (4) \cite{Bertram2007AA}; (5) \cite{Scoville2003ApJ}.
}
\end{table*}

\begin{longrotatetable}
\begin{deluxetable}{c c c c c c c c c c c c c c r@{$\pm$}l}
\tablecaption{Literature Sample \label{tab:lite}}
\tabletypesize{\scriptsize}
\tablehead{
\colhead{Object} &
\colhead{$z$} &
\colhead{$\log\,\lambda L_\lambda$(5100 \AA)} &
\colhead{$\log\,M_\mathrm{BH}$} &
\colhead{$\log\,M_*$} &
\colhead{$q$} &
\colhead{Ref.} &
\colhead{$S_\mathrm{CO}\Delta \nu$} &
\colhead{$\log\,L^\prime_\mathrm{CO}$} &
\colhead{$\log\,M_\mathrm{H2}$} &
\colhead{$W_{50}$} &
\colhead{$W_{20}$} &
\colhead{Ref.} &
\colhead{$\log\,L_\mathrm{IR}$} &
\mcl{2}{c}{$\log\,M_\mathrm{gas}$}\\
\colhead{} &
\colhead{} &
\colhead{$(\mathrm{erg\,s^{-1}})$} &
\colhead{$(M_\odot)$} &
\colhead{$(M_\odot)$} &
\colhead{} &
\colhead{} &
\colhead{$(\mathrm{Jy\,km\,s^{-1}})$} &
\colhead{$(\mathrm{K\,km\,s^{-1}\,pc^{2}})$} &
\colhead{$(M_\odot)$} &
\colhead{$(\mathrm{km\,s^{-1}})$} &
\colhead{$(\mathrm{km\,s^{-1}})$} &
\colhead{} &
\colhead{$\mathrm{(erg\,s^{-1})}$} &
\mcl{2}{c}{$(M_\odot)$} \\
\colhead{(1)} &
\colhead{(2)} &
\colhead{(3)} &
\colhead{(4)} &
\colhead{(5)} &
\colhead{(6)} &
\colhead{(7)} &
\colhead{(8)} &
\colhead{(9)} &
\colhead{(10)} &
\colhead{(11)} &
\colhead{(12)} &
\colhead{(13)} &
\colhead{(14)} &
\mcl{2}{c}{(15)}
}
\startdata
PG 0052+251 & 0.155 & 45.00 & 8.99 & \ph{$>$}11.05 &    0.55 &       2 & \ph{0$<$}2.0\ph{0$<$} &  \ph{0$<$}9.39\ph{0$<$} &  \ph{0$<$}9.88\ph{0$<$} & 429                  & 500\tnm{a}\phtf    & 3 & $44.51^{+0.02}_{-0.02}$ & 10.29 & 0.20         \\
PG 0157+001 & 0.164 & 44.95 & 8.31 & \ph{$>$}11.53 &    0.60 &       1 &  5.5$\pm$0.5          &  9.88$\pm$0.04          & 10.37$\pm$0.30          & 270\tnm{a}\phtf      & 315                & 4 & $45.85^{+0.03}_{-0.05}$ & 10.80 & 0.20         \\
PG 0804+761 & 0.100 & 45.03 & 8.55 & \ph{$>$}10.64 &    0.65 &       2 &  2.0$\pm$0.5          &  9.00$\pm$0.11          &  9.49$\pm$0.32          & 755                  & 881\tnm{a}\phtf    & 5 & $43.83^{+0.07}_{-0.05}$ &  8.79 & 0.21         \\
PG 0838+770 & 0.131 & 44.70 & 8.29 & \ph{$>$}11.14 & \nodata & \nodata &  2.5$\pm$0.4          &  9.34$\pm$0.07          &  9.83$\pm$0.31          & \ph{0}60\tnm{a}\phtf & \ph{0}70           & 4 & $44.72^{+0.03}_{-0.04}$ & 10.21 & 0.20         \\
PG 0844+349 & 0.064 & 44.46 & 8.03 & \ph{$>$}10.69 &    0.39 &       1 & \ph{0}$<$1.5\ph{0$<$} &  \ph{0}$<$8.48\ph{0$<$} &  \ph{0}$<$8.97\ph{0$<$} & \nodata              & \nodata            & 5 & $43.61^{+0.02}_{-0.02}$ & 10.01 & 0.21         \\
PG 1202+281 & 0.165 & 44.57 & 8.74 & \ph{$>$}10.86 &    0.92 &       2 & \ph{0}$<$2.4\ph{0$<$} &  \ph{0}$<$9.53\ph{0$<$} &       $<$10.02\ph{0$<$} & \nodata              & \nodata            & 6 & $44.53^{+0.03}_{-0.03}$ &  9.51 & 0.20         \\
PG 1226+023 & 0.158 & 45.99 & 9.18 & \ph{$>$}11.51 &    0.65 &       2 &  1.82$\pm$0.02        &  9.37$\pm$0.01          &  9.86$\pm$0.30          & 490                  & 572                & 8 & $42.58^{+0.47}_{-0.57}$ &  9.11 & 0.57         \\
PG 1309+355 & 0.184 & 44.98 & 8.48 & \ph{$>$}11.22 & \nodata &       1 & \ph{0}$<$0.6\ph{0$<$} &  \ph{0}$<$9.02\ph{0$<$} &  \ph{0}$<$9.51\ph{0$<$} & \nodata              & \nodata            & 3 & $44.41^{+0.04}_{-0.04}$ & 10.40 & 0.23         \\
PG 1351+640 & 0.087 & 44.81 & 8.97 & \ph{$>$}10.63 &    0.98 &       2 &  2.7$\pm$0.5          &  9.01$\pm$0.08          &  9.50$\pm$0.31          & 260\tnm{a}\phtf      & 303                & 4 & $44.78^{+0.04}_{-0.05}$ &  9.67 & 0.20         \\
PG 1402+261 & 0.164 & 44.95 & 8.08 & \ph{$>$}10.86 &    0.45 &       1 & \ph{0$<$}2.0\ph{0$<$} &  \ph{0$<$}9.44\ph{0$<$} &  \ph{0$<$}9.93\ph{0$<$} & \nodata              & \nodata            & 3 & $45.01^{+0.04}_{-0.04}$ &  9.93 & 0.20         \\
PG 1411+442 & 0.089 & 44.60 & 8.20 & \ph{$>$}10.84 &    0.71 &       1 & \ph{0}$<$1.8\ph{0$<$} &  \ph{0}$<$8.85\ph{0$<$} &  \ph{0}$<$9.34\ph{0$<$} & \nodata              & \nodata            & 5 & $44.14^{+0.03}_{-0.03}$ &  9.90 & 0.21         \\
PG 1415+451 & 0.114 & 44.53 & 8.14 &      $>$10.53 & \nodata & \nodata &  2.1$\pm$0.3          &  9.14$\pm$0.06          &  9.63$\pm$0.31          & \ph{0}90\tnm{a}\phtf & 105                & 4 & $44.40^{+0.02}_{-0.01}$ &  9.73 & 0.20         \\
PG 1440+356 & 0.077 & 44.52 & 7.60 & \ph{$>$}11.05 &    0.66 &       1 &  6.6$\pm$0.6          &  9.29$\pm$0.04          &  9.78$\pm$0.30          & 310\tnm{a}\phtf      & 362                & 4 & $44.77^{+0.02}_{-0.01}$ &  9.95 & 0.20         \\
PG 1444+407 & 0.267 & 45.17 & 8.44 & \ph{$>$}11.15 &    0.78 &       1 & \ph{0$<$}0.7\ph{0$<$} &  \ph{0$<$}9.42\ph{0$<$} &  \ph{0$<$}9.91\ph{0$<$} & 257                  & 300\tnm{a}\phtf    & 3 & $44.97^{+0.05}_{-0.05}$ &  9.51 & 0.24         \\
PG 1545+210 & 0.266 & 45.40 & 9.47 & \ph{$>$}11.15 & \nodata &       1 & \ph{0}$<$1.0\ph{0$<$} &  \ph{0}$<$9.57\ph{0$<$} &       $<$10.07\ph{0$<$} & \nodata              & \nodata            & 3 & $<$44.03                & \mcl{2}{c}{$<$10.38} \\
PG 1613+658 & 0.129 & 44.81 & 9.32 & \ph{$>$}11.46 & \nodata &       1 &  8.0$\pm$0.6          &  9.83$\pm$0.03          & 10.32$\pm$0.30          & 400\tnm{a}\phtf      & 467                & 4 & $45.39^{+0.02}_{-0.02}$ & 10.56 & 0.20         \\
PG 1700+518 & 0.282 & 45.69 & 8.61 & \ph{$>$}11.39 &    0.49 &       1 &  3.9$\pm$0.7          & 10.22$\pm$0.08          & 10.71$\pm$0.31          & 260\tnm{a}\phtf      & 303                & 7 & $45.81^{+0.02}_{-0.05}$ & 10.64 & 0.20         \\
\enddata
\tablenotetext{a}{The line width was originally provided in the literature.}
\tablecomments{
(1) Source name.
(2) Redshift.
(3) AGN monochromatic luminosity of the continuum at 5100 \AA.
(4) BH mass.
(5) Stellar mass of the host galaxy.  The uncertainty of the direct
measurements is $\sim 0.3$ dex.  The lower limits come from bulge masses
estimated from the BH mass using the $M_\mathrm{BH}-M_\mathrm{bulge}$ relation.
See Table 1 of \citealt{Shangguan2018ApJ} for more details.
(6) Axial ratio, derived from GALFIT modeling of the quasar host galaxies.
(7) References of the axial ratio.
(8) Integrated line flux of CO(1--0) emission.
(9) CO(1--0) line luminosity.
(10) Molecular gas mass derived from CO line luminosity, assuming
$\alpha_\mathrm{CO}=3.1\,M_\odot\,\mathrm{(K\,km\,s^{-1}\,pc^{2})^{-1}}$.
(11) The width of the CO integrated profile at 50 percent of its maximum.
(12) The width of the CO integrated profile at 20 percent of its maximum.
The flagged line widths were originally provided in the literature; other values  converted assuming $W_{20}/W_{50}=1.17$.
(13) References for the CO(1--0) measurements.
(14) IR luminosity of the host galaxy from spectral energy distribution decomposition by \cite{Shangguan2018ApJ}.
(15) Total gas mass derived from the dust mass.
Columns (2)--(5) and (15) are collected from Table 1 of \cite{Shangguan2018ApJ}.
References: (1) \cite{Kim2017ApJS}; (2) Y. Zhao et al. (in preparation);
(3) \cite{Casoli2001ASPC}; (4) \cite{Evans2006AJ}; (5) \cite{Scoville2003ApJ};
(6) \cite{Evans2001AJ}; (7) \cite{Evans2009AJ}; (8) \cite{Husemann2019ApJ}.
}
\end{deluxetable}
\end{longrotatetable}

\end{document}